\documentclass[structabstract]{aa}  
\usepackage{graphicx}
\usepackage{txfonts}
\usepackage[authoryear]{natbib}
\usepackage{subfig}
\begin{document}
   \title{An interferometric study of the post-AGB binary 89\,Herculis}
   \subtitle{I Spatially resolving the continuum circumstellar environment at optical and near-IR wavelengths with the VLTI, NPOI, IOTA, PTI, and the CHARA Array
   \thanks{Based on observations made with ESO Telescopes at the La Silla Paranal Observatory under program ID 079.D-0013 and 089.D-0576.}\fnmsep
   \thanks{FITS files of the calibrated visibilities are only available at the CDS via anonymous ftp to cdsarc.u-strasbg.fr (130.79.128.5) or via
   {http://cdsweb.u-strasbg.fr/cgi-bin/qcat?J/A+A/}.}}
   
   \author{M. Hillen\inst{1}
          \and
          T. Verhoelst\inst{1,2}
          \and
          H. Van Winckel\inst{1} 
          \and
          O. Chesneau\inst{3}
          \and
          C. A. Hummel\inst{4}
          \and
          J. D. Monnier\inst{5}
          \and
          C. Farrington\inst{6}
          \and
          C. Tycner\inst{7}
          \and
          D. Mourard\inst{3}
          \and
          T. ten Brummelaar\inst{6}
          \and
          D. P. K. Banerjee\inst{8}
          \and
          R. T. Zavala\inst{9}
          }

   \institute{Instituut voor Sterrenkunde (IvS), KU Leuven,
              Celestijnenlaan 200D, B-3001 Leuven, Belgium\\
              \email{Michel.Hillen@ster.kuleuven.be}
          \and
             Belgian Institute for Space Aeronomy, Brussels, Belgium      
          \and 
             Laboratoire Lagrange, UMR7293, Univ. Nice Sophia-Antipolis, CNRS, Observatoire de la C\^ote d'Azur, 06300 Nice, France
          \and
             European Southern Observatory, Karl-Schwarzschild-Str. 2, 85748 Garching
          \and
             University of Michigan, 941 Dennison Building, 500 Church Street, Ann Arbor, MI 48109-1090
          \and
             The CHARA Array of Georgia State University, Mt. Wilson Observatory, Mt. Wilson, CA 91023
          \and
             Department of Physics, Central Michigan University, Mt. Pleasant, MI 48859
          \and
             Physical Research Laboratory, Navrangpura, Ahmedabad, India 380009
          \and
             U.S. Naval Observatory, Flagstaff Station, 10391 W. Naval Obs. Rd., Flagstaff, AZ 86001, USA
             }

   \date{Received 01 April, 2013; accepted 25 July, 2013}
   \authorrunning{Hillen et al.}
   \titlerunning{}

% \abstract{}{}{}{}{} 
% 5 {} token are mandatory
 
  \abstract
   {Binary post asymptotic giant branch (post-AGB) stars are interesting laboratories to study both the evolution of binaries as well as the structure of circumstellar disks. }
   {A multiwavelength high angular resolution study of the prototypical object 89\,Herculis is performed with the aim of identifying and locating the different emission
   components seen in the spectral energy distribution.}
   {A large interferometric data set, collected over the past decade and covering optical and near-infrared (near-IR) wavelengths, is analyzed in combination with the spectral
   energy distribution (SED) and 
   flux-calibrated optical spectra. In this first paper only simple geometric models are applied to fit the interferometric data. 
   Combining the interferometric constraints with the photometry and the optical spectra, we re-assess the energy budget of the post-AGB star and 
   its circumstellar environment.}
   {We report the first (direct) detection of a large (35-40\%) optical circumstellar flux contribution and spatially resolve its emission region. Given this 
   large amount of reprocessed and/or redistributed optical light, the fitted size of the emission region is rather compact and fits with(in) 
   the inner rim of the circumbinary dust disk. This rim dominates our K band data through thermal emission and is rather compact, 
   emitting significantly already at a radius of twice the orbital separation. We interpret the circumstellar optical flux as 
   due to a scattering process, with the scatterers located in the extremely puffed-up inner rim of the disk and possibly also in a bipolar outflow 
   seen pole-on. A non local thermodynamic equilibrium (non-LTE) gaseous origin in an inner disk cannot be excluded but is considered highly unlikely.}
   {This direct detection of a significant amount of circumbinary light at optical wavelengths poses several significant questions regarding our understanding of both
   post-AGB binaries and the physics in their circumbinary disks. Although the identification of the source of emission/scattering remains inconclusive 
   without further study on this and similar objects, the implications are manifold.}


   \maketitle
%
%________________________________________________________________

\section{Introduction}
\citet{1993AAWaters} defined 89\,Her as the prototype of a new class of post-AGB binaries surrounded by circumbinary dust disks. Subsequent studies 
\citep{2006AAdeRuyter,2009AAvanWinckel} found a link between binarity and the presence of both hot and cool circumstellar material 
for many high galactic latitude supergiants. The infrared excesses are very similar to those observed for young stellar 
objects (YSOs) and strongly suggest a disk-like origin in all these sources \citep[][and references therein]{2011AAGielen}.
%The combination of a high galactic latitude with a low surface gravity can be associated more easily with an evolved low-mass star than a Pop. I supergiant. 

The orbital elements of post-AGB binaries show their companion stars to be most likely on the main sequence
and to have separations of $\sim$1~AU \citep{2009AAvanWinckel}, which are too small to harbor an AGB star. Moreover, most post-AGB binaries are 
still O-rich \citep{2011AAGielen}, which suggests that their evolution was cut short during a very short-lived phase. In this phase, 
a significant part of the primary's mass is expelled into a circumbinary orbit 
while avoiding a common envelope situation \citep[see, e.g.,][]{1998ApJMastrodemos,2003ARAAvanWinckel,2007BaltAFrankowski}. 
Due to the large and badly constrained distances, luminosities are not well known; this hampers 
progress in connecting them to other classes of the binary zoo. But disk formation is found to be a mainstream process around
evolved stars, as was recently found in the Large Magellanic Cloud \citep[LMC][]{2011AAvanAarle}.
%Many questions remain such as the link with other evolved binaries and whether this scenario is compatible with the observed e-log P diagram \citep{2009AAJorissen,2013AADermine}.

%The peculiar photospheric abundances \citep{2005ApJGiridhar,2007AAReyniers,2009AAGielen} 
%are most easily interpreted as due to selective re-accretion of H-rich material \citep{1992AAWaters}, i.e. gas cleaned from refractory elements through 
%radiation pressure on condensed dust. This concept is well-established, 
%but was never directly confirmed due to the small angular scales involved. 
With optical interferometers now online,
it is possible to resolve the prevailing circumstellar geometries for stars all over the Hertzsprung-Russell diagram, from optical to mid-IR wavelengths. 
For YSOs this has led to the discovery of an empirical size-luminosity relation 
\citep{2001ApJMillanGabet,2002ApJMonnier}, which in turn has led to the current paradigm \citep{2010ARAADullemond} of a passive dusty disk 
with an optically thin cavity and the inner radius set by the dust sublimation distance. The appearance and evolution of YSOs and 
binary post-AGB disks are likely dictated by the same processes, but differences in gravity, luminosity,
and evolutionary timescales should lead to different disk evolution.
Disks around binary post-AGB stars have been much less the subject of interferometric study. A few objects were investigated \citep[][]{2006AADeroo,2007AADeroo} 
with the MIDI and AMBER instruments on the Very Large Telescope Interferometer (VLTI), providing the final piece of evidence that the IR excess arises 
from a very compact disk-like geometry. But a detailed analysis of the compact structures remains to be performed.

%A hot topic in planetary nebulae (PNe) research is whether the detection of a large fraction of asymmetric 
%(often bipolar) PNe is predominantly caused by the presence of a binary in the center 
%\citep{2009PASPDeMarco}. Recently, detections of both puffed-up \citep{2007AAChesneau,2011AALykou} and 
%flat \citep{2007ApJSu,2011AJChu,2012ApJSBilikova} compact disks in PNe have been reported,
%suggesting there is an evolutionary link with the post-AGB disk sources. Due to the large uncertainty with respect 
%to the binary interaction channels that can lead to disk formation, it is unclear whether the different kinds of disks are linked.

We present in this paper the first detailed multiwavelength interferometric analysis of the prototypical object 89\,Herculis. 
\citet{2007AABujarrabal} presented some N-band interferometric data in addition to their Plateau de Bure Interferometer CO maps. 
The latter contained two nebular components, 
1) an extended hour-glass-like structure with expansion velocities of 7~km~s$^{-1}$ 
and 2) the unresolved, smaller than 0.4$\arcsec$, Keplerian disk with a velocity dispersion of only 5~km~s$^{-1}$. 
They also derived the inclination of the system under the assumption that the symmetry axis of the resolved 
outflow is perpendicular to the binary orbital plane (Table~\ref{table:basicpars}). 
Attempts to spatially resolve any extended structure at optical and mid-IR wavelengths with the Hubble Space Telescope (HST) 
\citep{2008ApJSiodmiak} and VISIR/VLT \citep{2011MNRASLagadec} failed to detect any emission at angular scales
beyond $\sim$50 and $\sim$300~mas, respectively.

%In our attempt to spatially resolve the physical components seen in the H$\alpha$ line with the VEGA instrument 
%on the CHARA Array, the optical continuum visibilities were surprisingly low (see further). 
We collected a large set of 
observations from the VLTI, the Navy Precision Optical Interferometer (NPOI), the Palomar Testbed Interferometer (PTI), the Infrared Optical Telescope Array (IOTA) 
and the Center for High Angular Resolution Astronomy (CHARA) Array, covering 0.5~to 2.2~$\mu$m and with
baselines from 15 to 278~m. Here we describe these observations and our simple geometric analysis. 
With these results, the SED is then redefined. In Paper II, a detailed radiative transfer modeling will be attempted to 
reproduce our observables.

\begin{table}
\caption{Stellar and binary parameters of 89\,Herculis.}             % title of Table
\label{table:basicpars}      % is used to refer this table in the text
\centering                          % used for centering table
\begin{tabular}{c c c c}        % centered columns (4 columns)
\hline\hline                 % inserts double horizontal lines
Parameter & Value & Error & References \\    % table heading 
\hline                        % inserts single horizontal line
   Sp.T. & F2Ibe & - & -  \\
   Teff (K) & 6550 & 100 & 1,2 \\      % inserting body of the table
   log g & 0.55 & 0.25 & 1,2 \\
   $[\mathrm{Fe}/\mathrm{H}]$ & -0.5 & 0.2 & 1,2 \\   
   P$_{\mathrm{orb}}$ (d) & 288.36 & 0.71 & 3 \\
   e & 0.189 & 0.074 & 3 \\
   a$_1 \sin i$ (AU) & 0.080 & 0.007 & 3 \\
   f(m) (M$_\odot$) & 0.00084 & 0.00022 & 3 \\
   i($^\circ$) & 12 & 3 & 4 \\
   $\pi$ (mas) & 0.76 & 0.23 & 5 \\
   d (kpc) & 1.5 & $^{+1.0}_{-0.5}$ & 5 \\
\hline                                   %inserts single line
\end{tabular}
\tablebib{(1) \citet{1990ApJLuck}, (2) \citet{2011BaltAKipper}; (3) \citet{1993AAWaters}; (4) \citet{2007AABujarrabal}; (5) \citet{2007AAvanLeeuwen}}
\end{table}

%__________________________________________________________________

\section{Observations and data reduction}
\subsection{Interferometry}
\begin{figure*}
\centering
   \includegraphics[width=8cm,height=7cm]{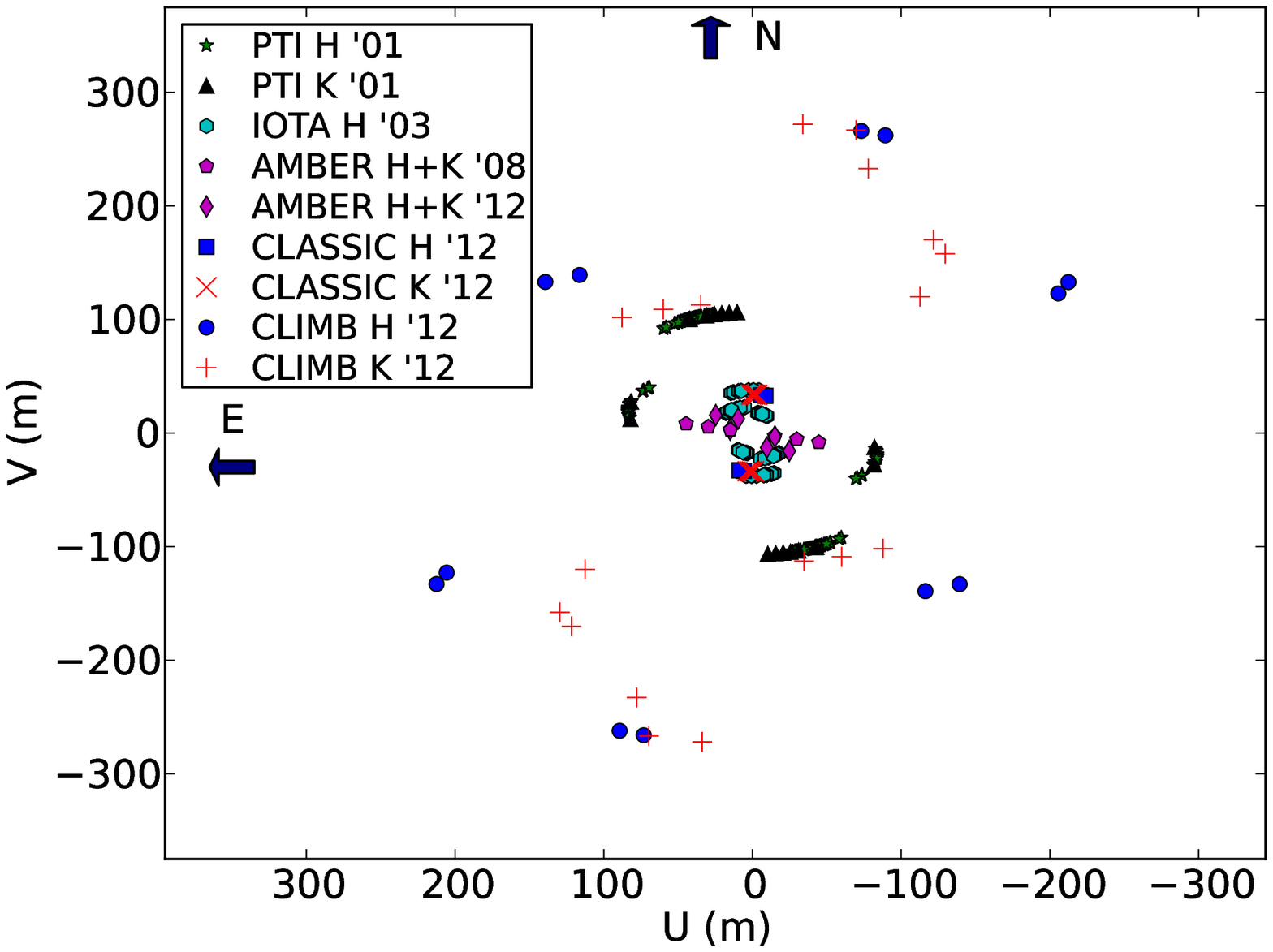} 
   \hspace{1cm}
   \includegraphics[width=8cm,height=7cm]{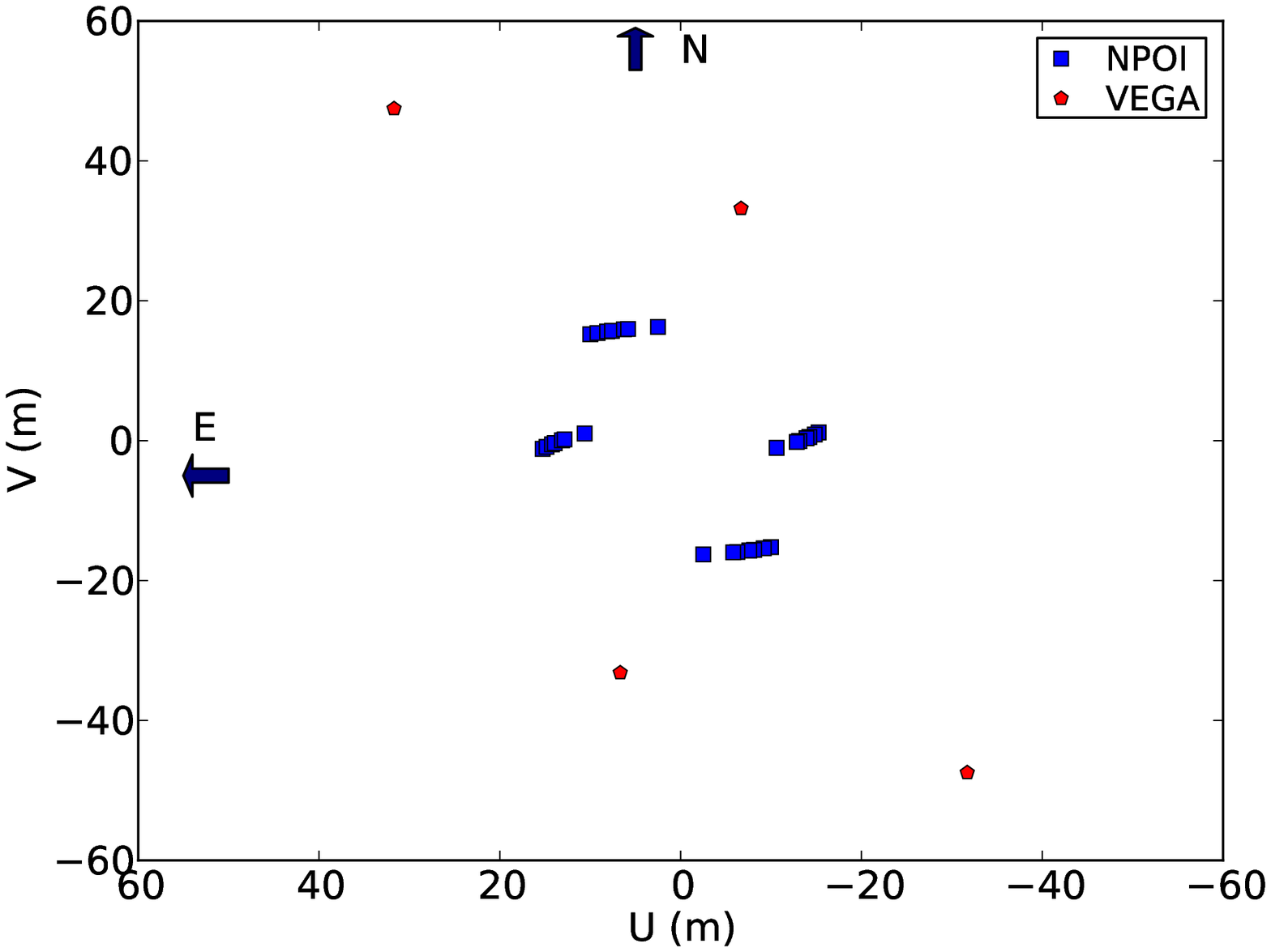} 
     \caption{Baseline coverage of our observations. Left panel: the total near-IR UV coverage, right panel: total optical UV coverage.}
     \label{figure:uvcoverage}
\end{figure*}

\begin{figure*}
\centering
 %\sidecaption
   %\includegraphics[width=8cm,height=7cm]{89Her_geometric_Hv2.eps} %{RingdiskFstar055thetastar045inner1width65AMBER0812.eps} 
   \includegraphics[width=8cm,height=7cm]{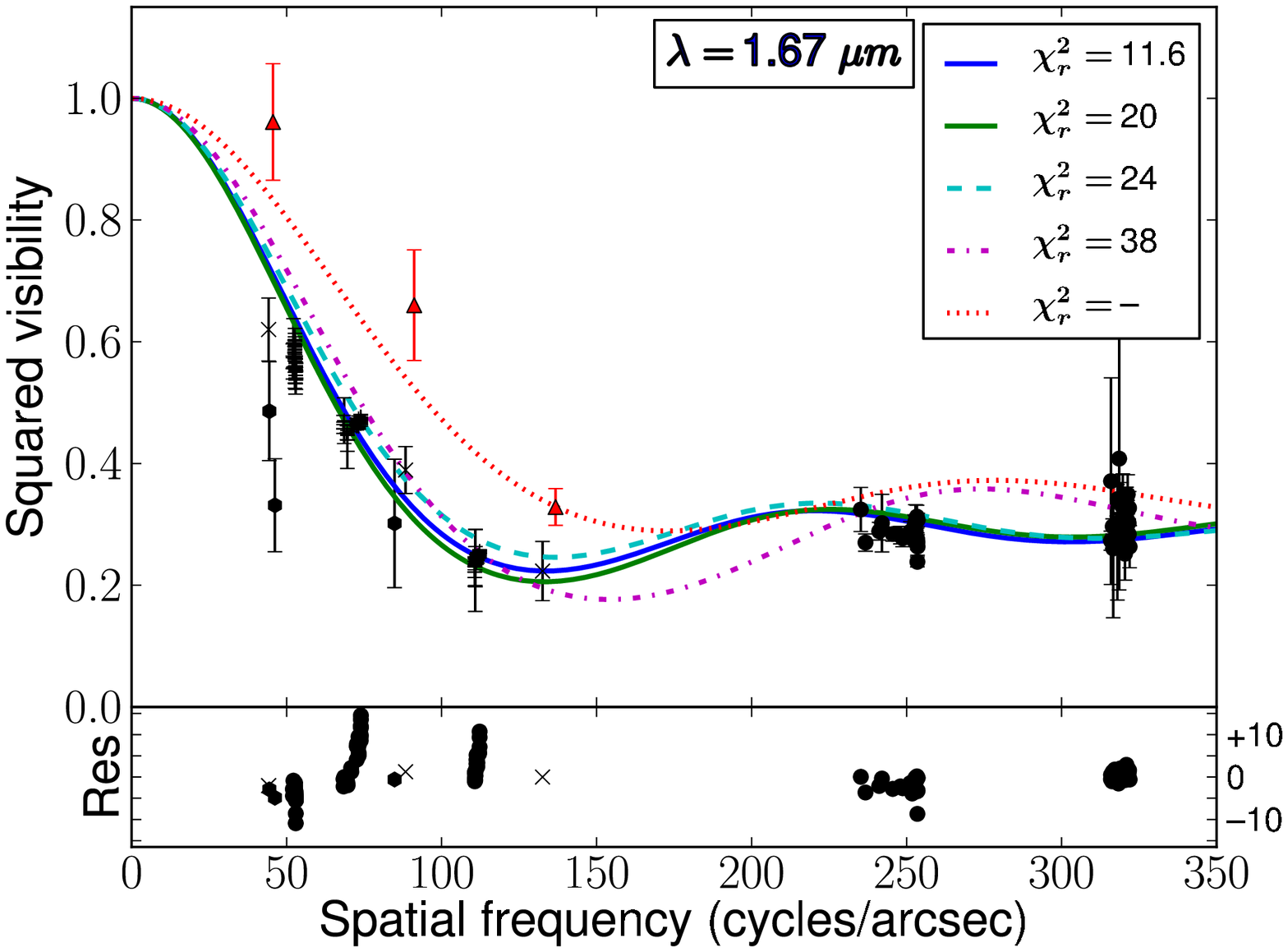} %{RingdiskFstar055thetastar045inner1width65AMBER0812.eps} 
   \hspace{1cm}
   \includegraphics[width=8cm,height=7cm]{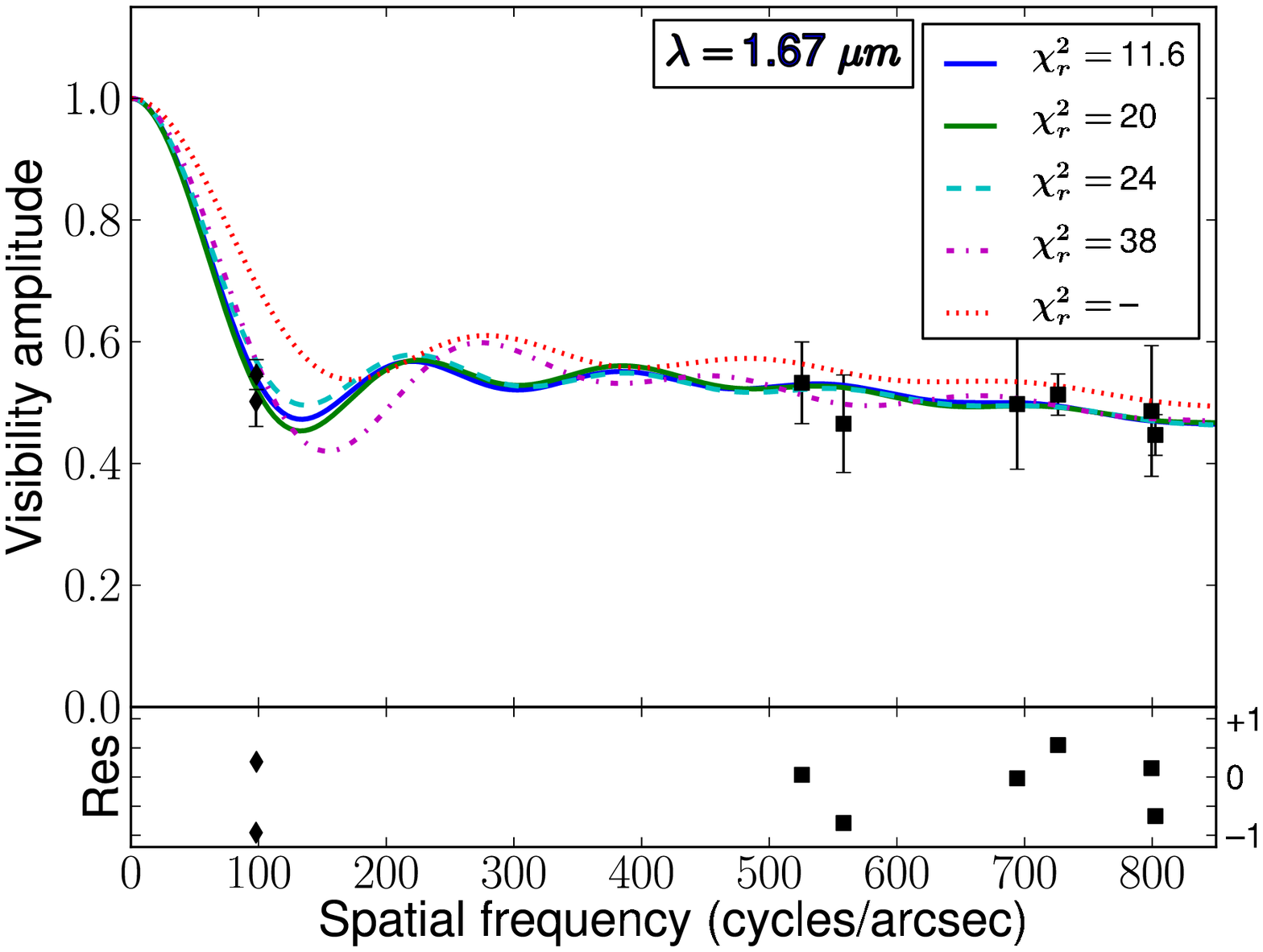} %{RingdiskFstar055thetastar045inner1width65CHARA.eps}
     %\caption{Best-fit star+ring model in H band. Left: the model with the AMBER squared visibilities. Right: the model with the CHARA visibility amplitudes.}
     \caption{Left: the AMBER (crosses: 2008; pentagons: 2012), PTI (circles) and IOTA (plusses) H band squared visibilities. Right: the CHARA Array CLASSIC (diamonds) and 
     CLIMB (squares) H band visibility amplitudes. The red triangles are the 2007 AMBER data, which are treated separately. In both plots the same 
     five geometric star+ring models are depicted. The full blue and green, dashed cyan, and dot-dashed magenta lines are models with inner diameters 
     and widths equal to 3.0 and 4.6, 4.1 and 4.0, 1.0 and 5.5, and 5.0 and 2.5~mas, respectively. The dotted red line is the best-fit model to 
     the 2007 AMBER data (see text). The legend denotes the $\chi^2_r$ value of the corresponding model, except for the red one, 
     which formally has too few data points.
     }
     \label{figure:Hsinglering}
\end{figure*}

A log of all interferometric observations is presented in Table~\ref{table:obslog}. Below we discuss the individual data sets. 
Figure~\ref{figure:uvcoverage} shows the obtained UV coverage. Figs.~\ref{figure:AMBERAllvis},~\ref{figure:Hsinglering},~\ref{figure:Ksinglering},
and~\ref{figure:closures+optical} (some only available online) depict the near-IR calibrated visibilities and the IOTA closure phases. 
The optical visibilities are visible in Figs.~\ref{figure:closures+optical} and~\ref{figure:allOptVis}.

% Figs.~\ref{figure:Hsinglering} and~\ref{figure:closures+optical}, 
% and Figs.~\ref{figure:AMBERAllvis} and~\ref{figure:Ksinglering} in the online appendix depict the near-IR calibrated visibilities and IOTA closure phases. 
% The optical visibilities are visualized in Figs.~\ref{figure:closures+optical}, and Fig.~\ref{figure:allOptVis} in the online appendix.

\begin{table*}
\caption{Observing log of our interferometric observations.}             % title of Table
\label{table:obslog}      % is used to refer this table in the text
\centering                          % used for centering table
\begin{tabular}{c c c c c c c c}        % centered columns (4 columns)
\hline\hline                 % inserts double horizontal lines
Date & Mean MJD & Instrument & Stations & Nr. Obs. & $\lambda$($\mu$m) & $\delta \lambda$ ($\mu$m) & Calibrators\tablefootmark{a} \\    % table heading 
\hline                        % inserts single horizontal line
   2001 May 8 & 52037.5 & PTI & NW & 8 & 1.65 & 0.05 &  HD166014, HD168914 \\
   2001 Jun 6 & 52066.4 & PTI & NW & 5 & 1.65 & 0.05 &  HD166014, HD168914 \\
   2001 Jun 23 & 52083.3 & PTI & NS & 7 & 2.1 & 0.1 & HD166014, HD168914  \\
   2001 Jul 2,15 & 52098.5 & PTI & NW & 7 & 1.65 & 0.05 &  HD166014, HD168914 \\
   2001 Jul 28 & 52118.2 & PTI & NW & 2 & 2.1 & 0.1 & HD166014, HD168914  \\
   2001 Aug 2,6,10,11,18 & 52132. & PTI & NS & 27 & 1.65 & 0.05 & HD166014, HD168914  \\
   2003 Jun 10 & 52800.3 & PTI & NS & 7 & 1.65 & 0.05 &  HD166014, HD168914 \\
   2003 Jun 15,16 & 52805.8 & IOTA & N35-C10-S15 & 23 & 1.65 & 0.25 & HD166014, HD168914 \\
   2007 Apr 13 & 54203.38 & AMBER & E0-G0-H0 & 1 & 1.50-2.50 & 0.03 & HD165524 \\
   2008 Mar 31 & 54556.39 & AMBER & E0-G0-H0 & 1 & 1.50-2.50 & 0.03 & HD165524  \\
   2011 Jun 8 &  55720.5 & VEGA & S1-S2 & 1 & 0.672 & 0.015 & HD168914 \\
   2011 Aug 24 & 55797.5 & VEGA & E1-E2 & 1 & 0.672 & 0.015 & HD168914 \\
   2011 Oct 9 & 55844.16 & NPOI & AC0-AE0 & 7 & 0.56-0.85 & 0.025 & HR6787 \\   
   2011 Oct 9 & 55844.16 & NPOI & AE0-AW0 & 7 & 0.56-0.85 & 0.025 & HR6787 \\   
   2012 May 1 & 56048.43 & CLASSIC & S1-S2 & 4 & 2.13 & 0.3 & HD162828, HD163948, HD164730 \\
   2012 May 1 & 56048.48 & CLASSIC & S1-S2 & 2 & 1.65 & 0.2 & HD161239, HD164730 \\      % inserting body of the table
   2012 June 27 & 56105.15 & AMBER & A1-C1-D0 & 1 & 1.50-2.50 & 0.03 & HD165524   \\
   2012 July 3 & 56111.46 & CLIMB & S2-E2-W2 & 1 & 2.13 & 0.3 & HD166230 \\
   2012 July 4 & 56112.41 & CLIMB & S1-E2-W2 & 2 & 2.13 & 0.3 & HD166230, HD168914 \\   
   2012 July 7 & 56115.43 & CLIMB & S1-E2-W1 & 2 & 1.65 & 0.2 & HD166230, HD168914 \\

\hline                                   %inserts single line
\end{tabular}
\tablefoot{\tablefoottext{a}{Calibrator UD diameters (mas): HD166014 = 0.58$\pm$0.03, HD168914(R) = 0.45$\pm$0.03, HD168914(HK) = 0.46$\pm$0.03,
HD165524 = 1.11$\pm$0.02 \citep{2005AAMerand}, HR6787 = 0.34$\pm$0.02, HD162828 = 0.72$\pm$0.05, 
HD161239 = 0.62$\pm$0.04, HD163948 = 0.52$\pm$0.03, HD164730 = 0.75$\pm$0.05, HD166230 = 0.41$\pm$0.03}}
\end{table*}

\subsubsection{Near-IR}
%At near-IR wavelengths, observations were collected with four instruments on three different interferometers, leading to a good coverage in baseline length and position angle.
%Here we describe the near-IR observations obtained with four interferometers. The calibrated visibilities are depicted in 
%Figs.~\ref{figure:Hsinglering} and~\ref{figure:closures+optical}, and Figs.~\ref{figure:AMBERAllvis} and~\ref{figure:Ksinglering} in the online appendix.

\paragraph{\textbf{VLTI}}
The object 89\,Herculis was observed with the AMBER instrument \citep{2007AAPetrov} on the VLTI using the Auxiliary Telescopes (ATs)
in 2007, 2008, and 2012, all in the low-spectral-resolution (R=30) mode covering the H and K bands with detector integration times (DITs) 
of 25 and 50~ms and without the use of the fringe tracker FINITO. Because 89\,Her is quite 
northern for the Paranal site, observing conditions were never excellent. In 2007 the observations occurred under good 
atmospheric conditions, with good seeing ($\sim0.8\arcsec$) and coherence time (2.5-3.0\,ms). In 2008 conditions were a bit worse ($\sim1.0\arcsec$,2.5\,ms) and in 
2012 they were bad ($>1.0\arcsec$, 2\,ms, and variable). 
The data were reduced with \textit{amdlib} v3.0.3, which is provided by the Jean-Marie Mariotti Center (JMMC) \citep{2007AATatulli,2009AAChelli} 
and calibrated with observations of HD165524 \citep{2005AAMerand}. 

The 10\% of the best signal-to-noise ratio (S/N) frames were used for visibility amplitude estimation, while 80\% of all frames were used to 
compute the closure phases. The exact values of these thresholds have little influence on the final numbers. A small correction 
to the wavelength table, in the form of a linear offset of -0.08, -0.08, and -0.07 $\mu$m for 2007, 2008, and 2012, respectively, was performed
to align the H-K discontinuity correctly. Although the final calibration only used the actual calibrator, 
we checked the consistency of the calibration and the stability of the transfer function by repeating the calibration with each calibrator observed during that night.
The generally small spread from this procedure was added to the final error budget to account for the lack of a second dedicated calibrator measurement 
in the 2007 and 2008 data set. Although the 2007 and 2008 data were observed at the same spatial frequencies, the visibilities are very different. 
The difference is larger in H band, but the short-wavelength end of the K band is also affected, showing that it is not just the 
absolute calibration in H band. The largest effect is observed for the shortest baselines. 
Based on our careful analysis, and the stability of the transfer function, we cannot find a reason why the visibilities 
are so different, and so the observed trend might be a real time-variable effect. Nevertheless, we decided not to use 
the 2007 data in our analysis because they are clearly discrepant from the 2008 data, which in turn are consistent with the 2003 IOTA as well as with the 
2012 VLTI and CHARA results.

\onlfig{2}{
\begin{figure}
\centering
   \includegraphics[width=8cm,height=7cm]{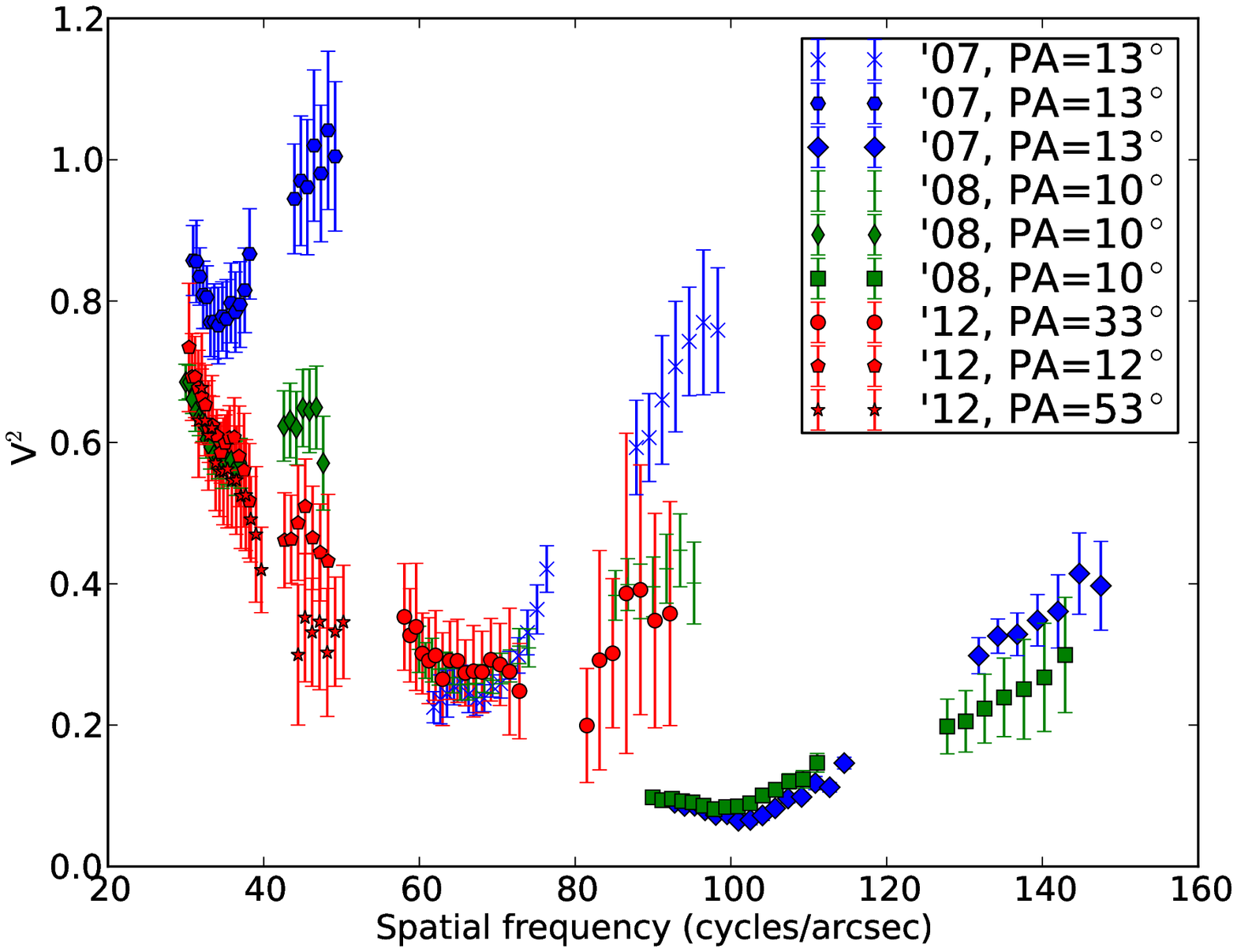} 
     \caption{The AMBER (LR-HK) squared visibilities as a function of the spatial frequency. The legend denotes the epoch and the position angle 
     (PA) of the baseline of the observation. All 2007 and 2008 data have a similar PA because the same co-linear baseline configuration 
     was used. Per baseline, the H and K band are separated by the HK-discontinuity, but still show a smooth 
     transition between the two.}
     \label{figure:AMBERAllvis}
\end{figure} 
}

\paragraph{\textbf{CHARA}}
Both the CLASSIC and CLIMB beam combiners on the Georgia State University CHARA Array interferometer were used to conduct 
observations in the H and K bands. The CHARA Array, located on Mount Wilson, is a six-telescope Y-shaped interferometric array and contains 
the longest optical/IR baselines currently in operation \citep{2005ApJtenBrummelaar}. Its 15 baselines, ranging from 34 to 331~m, 
provide resolutions up to $\sim$0.5 and $\sim$0.2~mas at near-IR and optical wavelengths respectively. 
In May 2012 observations were performed with the sensitive two-beam CLASSIC
beam combiner in both H and K band, using the S1S2 34\,m baseline. Three-beam CLIMB observations, in H and K band, were also obtained during three nights 
in July 2012 on three different station triplets with baseline lengths from 150 to 278~m. Conditions during all nights were good 
and stable. Data reduction and calibration were carried out with standard CHARA Array reduction software \citep{2012SPIEtenBrummelaar}. 
Calibrators were selected with the JMMC SearchCal %\footnote{http://www.jmmc.fr/searchcal/} 
tool \citep{2006AABonneau}. All calibrators 
are less than 7.5$^\circ$ away from the science target, have a brightness difference of at most 1~mag at the
wavelength of observation, and have visibility accuracies better than 7\% at the resolution of our observations. The final calibration uncertainties include 
both the scatter of the data from the repeatability of the measured calibrator visibility amplitude and a term accounting for the calibrator diameter uncertainties. 
The latter contribution is negligible for the CLASSIC data but becomes important for the long CLIMB baselines.

\paragraph{\textbf{PTI}}
We retrieved and calibrated observations from the PTI archive, which was made available by the NASA Exoplanet Science Institute
(NExScI)\footnote{http://nexsci.caltech.edu}. The PTI was a near-IR interferometer consisting of three 40~cm apertures at fixed 
separations of 85~to 110~m that were combined pairwise \citep{1999ApJColavita}. Observations were performed in four and five
spectral channels covering the H and K band, respectively. The data reduction was performed 
on site \citep{1999PASPColavita}, leaving only the calibration to be done with the \textit{nbCalib} 
tool \citep[provided by the NExScI,][]{1998SPIEBoden}. 89\,Her was observed during 30 nights between 1999 and 2003, 
equally spread over the H and K bands. The NW baseline was used during some nights in 2001, however, 
all other observations were taken with the NS baseline only. Four of the 30 nights were discarded due to inconsistent 
estimates of the visibility calibration when different calibration measurements were used. Since all observations at the same physical baseline 
give the same visibility within their (small) errors and probe the same uv coordinates, we select in the K band one particularly 
stable night per baseline to avoid too large a weight of the PTI data in the LITpro modeling (Sect.~\ref{section:geometric}).
All data are retained in the H band as their number matches that of the IOTA data set.
The two calibrators were originally selected using getCal, an SED-fitting routine maintained and distributed 
by the NExScI. However we fetched their angular diameters (which are well below the resolution limit of PTI) with SearchCal for HD168914 and 
from the catalog of \citet{2008ApJSVanBelle} for HD166014. Since SearchCal quotes a significantly smaller angular diameter 
($\sim$30\%) for the latter calibrator, we checked and confirmed the value found by \citet{2008ApJSVanBelle} by fitting Kurucz 
model atmospheres to archival photometry using the same tools as described in Sect.~\ref{section:stellarsed}. 
The final precision of the PTI observations is $\sim$5\%, which is typical for PTI and an object of the given brightness. 

\paragraph{\textbf{IOTA}}
Eight and 15 observations of 89\,Her were obtained at the IOTA interferometer \citep{2003SPIETraub} on 2003 June 15 and 16, respectively. 
This interferometer was located on Mt. Hopkins (AZ) and consisted of three 0.45~m telescopes that were movable about 17 stations along two orthogonal 
linear arms. Our observations were done at stations N35-C10-S15. They included three simultaneous 
baselines using the broadband H filter. The light beams from the 
three telescopes were interfered using the single-mode IONIC3 combiner \citep{2003SPIEBerger}. The data reduction was the same 
as described in previous IOTA papers \citep[e.g.,][]{2006ApJMonnier}. Also, closure phases
were procured with a precision of 1-3$^\circ$. The same calibrators were used as for the PTI.

\onlfig{4}{
\begin{figure*}
\centering
 %\sidecaption
   %\includegraphics[width=18cm]{CODEX_R52_LC-phase-2.2bb.eps} 
   %\includegraphics[width=8cm,height=7cm]{89Her_geometric_Kv2.eps} %{RingdiskFstar025thetastar045width6inner1AMBER0812.eps}
   \includegraphics[width=8cm,height=7cm]{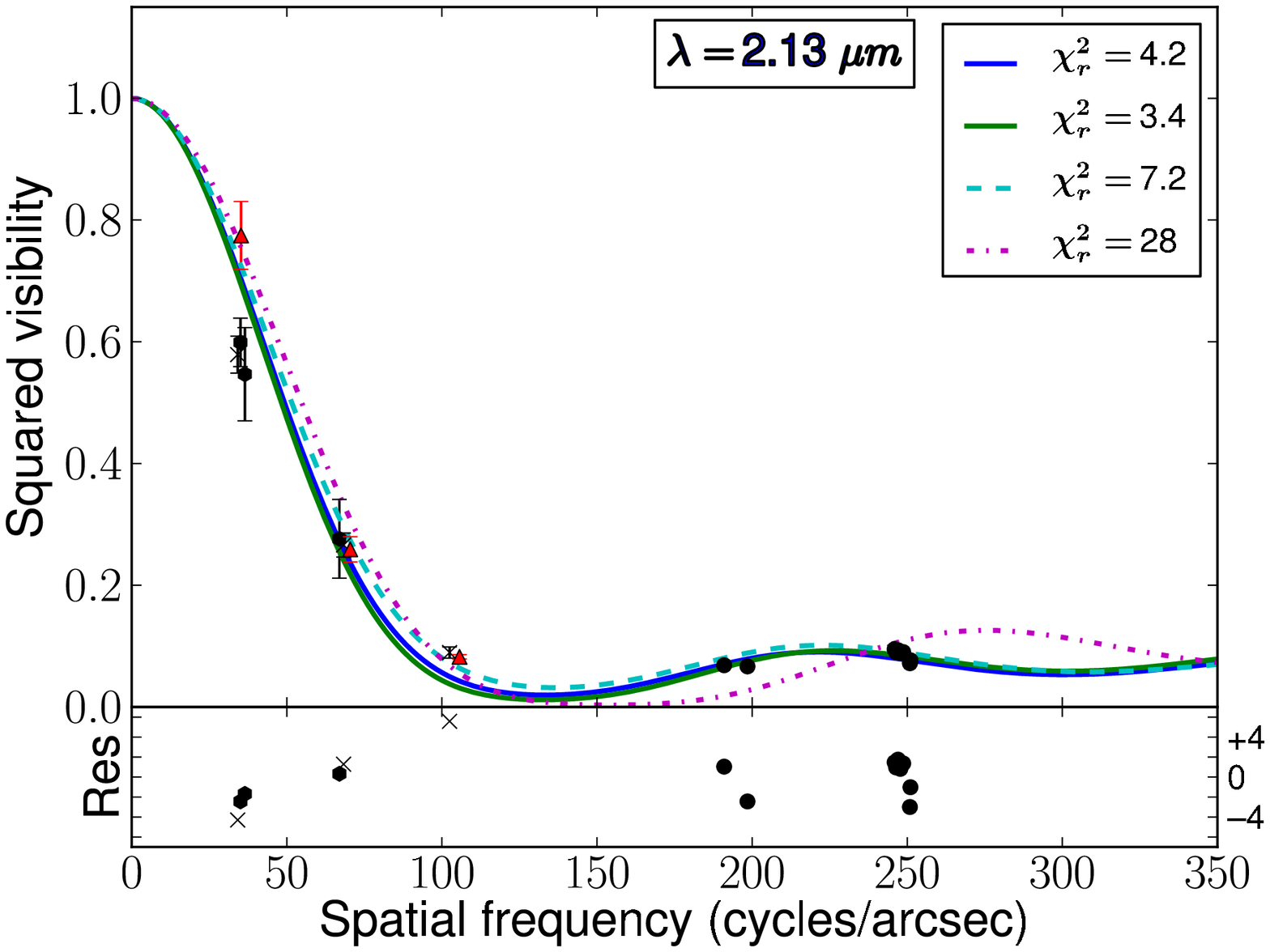} %{RingdiskFstar025thetastar045width6inner1AMBER0812.eps}
   \hspace{1cm}
   \includegraphics[width=8cm,height=7cm]{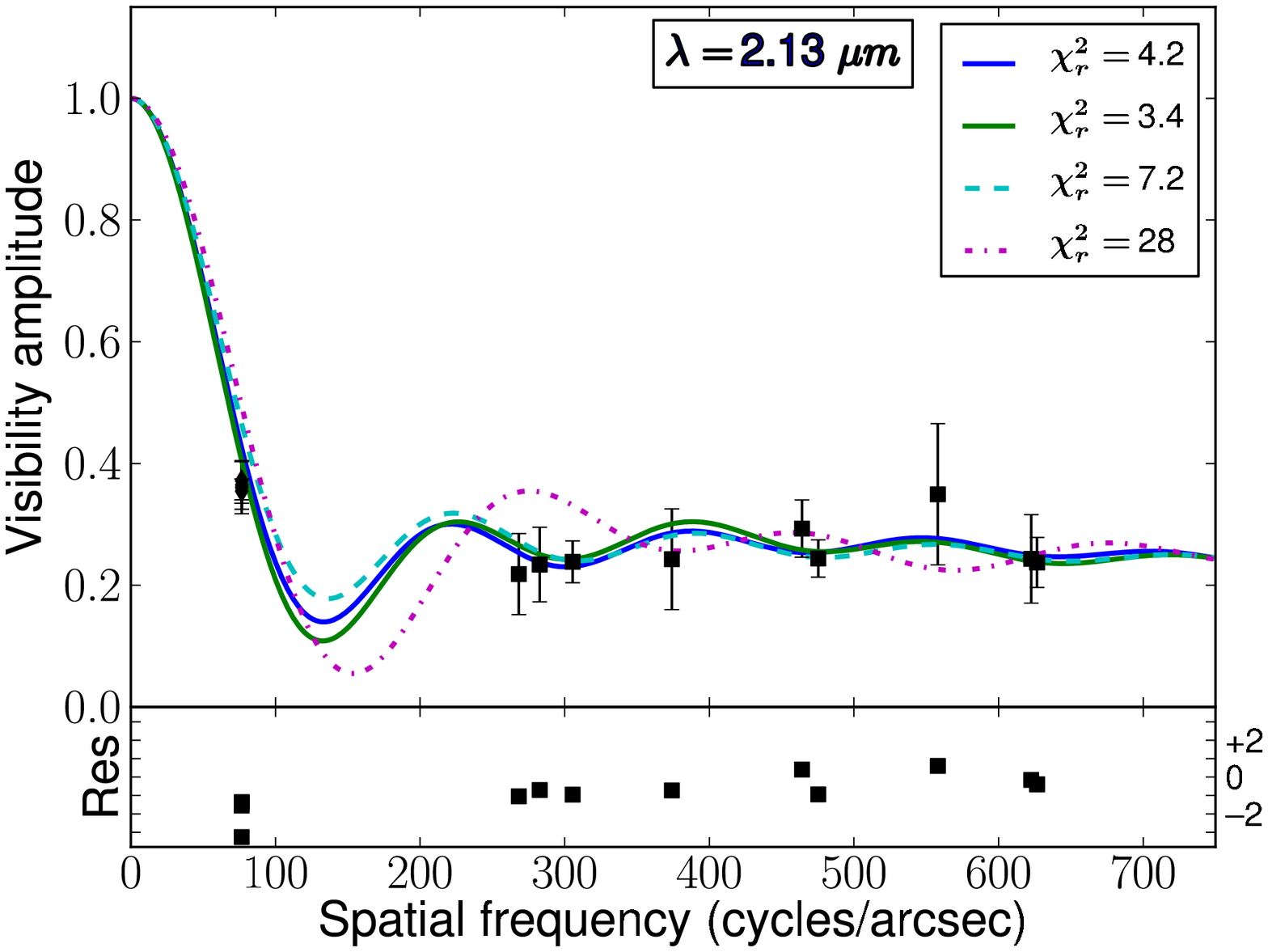} %{RingdiskFstar025thetastar045width6inner1CHARA.eps}
     %\caption{Best-fit star+ring model in K band. Left: the model with the AMBER squared visibilities. Right: the model with the CHARA visibility amplitudes.}
     \caption{The same as Fig.~\ref{figure:Hsinglering} but for K band (for which there are no IOTA data). The red model is not relevant in K and is not shown. }
     \label{figure:Ksinglering}
\end{figure*} 
}

\subsubsection{Optical}
%At optical wavelengths, observations were collected with the CHARA Array and the NPOI (see Fig.~\ref{figure:allOptVis} in the online appendix, and 
%Fig.~\ref{figure:closures+optical} for a visualization of the data).

\paragraph{\textbf{NPOI}}
The NPOI was described by
\citet{1998ApJArmstrong}. Observations of 89\,Her were carried
out in October 2011, but only in one night were the seeing conditions
good enough to allow further analysis of the data. For technical reasons,
only two baselines were available, between the astrometric east (AE)
and center (AC) station, and between the latter and the west station
(AW). Therefore, no closure phase data were recorded. Additional
``incoherent'' observations away from the fringe packet were recorded
for each star in order to determine the visibility bias. As a calibrator,
HR 6787 was interleaved with the seven observations of 89\,Her. The data
reduction followed the steps outlined by \citet{2003AJHummel}.
Based on apparent magnitudes and colors, the calibrator diameter was estimated to be 0.34 mas,
implying that it was unresolved by the
interferometer. Calibration uncertainties based on the repeatability
of the calibrator visibility amplitude ranged from a few percent in
the reddest channel (channel 1, 860 nm) up to about 10\% in the bluest
channel (channel 16, 560 nm).

\paragraph{\textbf{CHARA}}
Two observations were performed in June and August 2011 with the Visible spEctroGraph And Polarimeter \citep[VEGA instrument,][]{2009AAMourard} 
integrated within the CHARA Array. VEGA was used to recombine two telescopes, first S1S2 (34m) and then E1E2 (62m). 
The spectrograph is designed to sample the visible band from 0.45 to 0.85~$\mu$m, and VEGA is equipped with two photon-counting detectors
looking at two different spectral bands. The observations were performed with the medium-resolution setting of R=6000 under 
average-to-poor seeing conditions ($r_0\sim4-7$cm), with HD168914 as the calibrator.

The data reduction method is fully described in \citet{2009AAMourard} and we only briefly summarize it here. The spectra are extracted using a 
classical scheme of collapsing the 2D flux in one spectrum, wavelength calibration using a Th-Ar lamp, and normalization of the continuum by polynomial fitting.  
The raw squared visibilities were estimated by computing the ratio of the high-frequency energy to the low-frequency energy of the averaged spectral 
density. The same treatment was applied to the calibrators, whose angular diameter was estimated with SearchCal. 

The signal was extracted from a band of 15nm centered around 672.5nm. Given the limited S/N, 
it was not possible to extract a useful differential signal over the H$\alpha$ line.

\subsection{Photometry} \label{subsection:photometry}

Photometric measurements with reliable transmission curves and zero points were already analyzed
in \citet{2006AAdeRuyter} (i.e., GENEVA, IRAS, JCMT), but we also included AKARI data \citep{2007PASJMurakami}, UV photometry from TD1 \citep{1978csufThompson} and
from the ANS satellite \citep{1982AASWesselius}, JOHNSON 11-band measurements \citep{2002yCatDucati}, and Stromgren points from the GCPD \citep{1997AASMermilliod}. Also 
DIRBE photometry \citep{2004ApJSSmith} in the 3.5, 4.9, and 12~$\mu$m bands was included, as well as the WISE W4 band \citep{2012yCatCutri}. Other bands in the 
latter systems were excluded because of too large errors or
saturation effects. Near-IR JHK photometry of 89\,Her was obtained at a mean UT = 19.645 October 2012 with the Mount Abu 1.2m
telescope using the 256x256 HgCdTe NICMOS3 array of the Near-Infrared Imager/Spectrometer, which has a similar spectral response as the 2MASS system. 
The details of the observing and data reduction techniques are given in \citet{2002AABanerjee}. Calibration was done with the standard
star SAO 86043. %Finally, we also included  our Atacama Pathfinder Experiment (APEX) 850~$\mu$m measurement that 
%was obtained with the LABOCA camera \citep{2009AASiringo}. 

\begin{table*}
\caption{Observing log of the HERMES spectra that were used to obtain optical spectral slopes.}             % title of Table
\label{table:obslogHermes}      % is used to refer this table in the text
\centering                          % used for centering table
\begin{tabular}{c c c c c c c c c}        % centered columns (4 columns)
\hline\hline                 % inserts double horizontal lines
Date & MJD & Calibrator & V$_{\mathrm{cal}}$  & Sp. T. & T$_{\mathrm{exp,cal}}$ (s) & T$_{\mathrm{exp,sci}}$ (s) & Airmass$_{\mathrm{cal}}$ & Airmass$_{\mathrm{sci}}$ \\    % table heading 
\hline                        % inserts single horizontal line
   2011 Aug 05 & 55779.38 & HD152614 & 4.38 & B8V & 330 & 120 & 1.06 & 1.01 \\   
   2011 Sep 21 & 55826.35  & HD185395 & 4.48 & F4V & 360 & 150 & 1.09 & 1.07 \\
   2011 Oct 09 & 55844.32 & HD184006 & 3.77 & A5V & 190 & 120 & 1.09 & 1.11 \\      % inserting body of the table

\hline                                   %inserts single line
\end{tabular}
\end{table*}

\subsection{Spectroscopy}
Observations of 89\,Her were procured with the short (SWS) and long (LWS) wavelength spectrometers onboard ISO. We downloaded from the archive 
the reduced and calibrated spectrum, which covers 2.5-120~$\mu$m. 
The features in this spectrum were already analyzed by \citet{2002AAMolster}, so we refer to this paper for more information. The absolute calibration of the spectrum
agrees very well with the retrieved IRAS, AKARI, DIRBE, and WISE photometry, but flux levels at near-IR wavelengths seem to be slightly overestimated.

A near-IR spectrum was retrieved from the atlas of medium-resolution K band spectra published by \citet{1997ApJSWallace}. Despite the limited resolution and S/N, 
the spectrum shows the CO 2-0 and higher vibrational modes in emission, pointing at the presence of a high column density of hot gas. 

89\,Her is part of our spectroscopic monitoring program of evolved binaries \citep[see, e.g.,][]{2012AAVanWinckel,2012AAGorlova} 
with the HERMES spectrograph \citep[R=85000, 3770-9000\,$\mathring{A}$,][]{2011AARaskin} on the 1.2~m Mercator Telescope at the Roque de los Muchachos Observatory, La Palma. 
A detailed account of this large set of high-resolution spectra will
be presented elsewhere. Here we simply use some of them, i.e., those
with good S/N and close in time to the interferometric observations
and to suitable calibrator observations (Table~\ref{table:obslogHermes}), as an 
independent constraint on the optical continuum spectral slope. The latter can be calibrated well thanks to the high stability of the HERMES instrument, 
which is monitored by measuring standard stars each night. We use these standard star measurements for our a posteriori spectral shape calibration. 
The intrinsic  calibrator spectrum was determined by fitting a reddened Kurucz model atmosphere \citep{1993yCatKurucz}
to high-quality archival photometry (obtained from the same catalogs as described above, see Sect.~\ref{section:stellarsed} for a description of the fitting procedures). 
After convolution to a lower resolution wavelength grid, the intrinsic 
calibrator spectrum as well as the measured counts of both science and calibrator spectrum are normalized at 7400~$\mathring{A}$, where 
the HERMES throughput is maximal. The calibrated result is then the product of the intrinsic calibrator spectral 
shape with the measured science one, divided by the measured calibrator spectral shape. 
The scatter between calibrations with different calibrators and exposures is small.

%__________________________________________________________________

\section{A geometric analysis}\label{section:geometric}
Using the efficient LITpro software \citep{2008SPIETallonBosc} provided by the JMMC, we fit and analyze in this section the near-IR 
and optical interferometric data with simple geometric models to deduce sizes and flux ratios of the different
physical components. We start with the near-IR because of our experience with similar data on other disk sources 
and then continue with the unexplored optical regime. Given the nondetection
of large-scale nebulosity at optical and mid-IR wavelengths, there
is no need to take into account potential field-of-view (FOV) issues. 

\subsection{A ring in the near-IR}\label{subsection:nearIRring}
We first describe the general fitting process and then compare the results for the H and K bands. 
Our prime interest in this study is to constrain the broadband morphology of the object. 
We choose not to use the full spectral capabilities of AMBER (which even in the low resolution mode are
significantly better than for the other instruments) and PTI. For simplicity, we select the wavelength channels that are closest to the 
central wavelengths of the CLIMB, CLASSIC, and IONIC3 broad passbands.
Although a small bias might be introduced, the smooth and monotonic wavelength dependence of the visibility validates our approach.

\begin{figure*}
\centering
   \includegraphics[width=8cm,height=7cm]{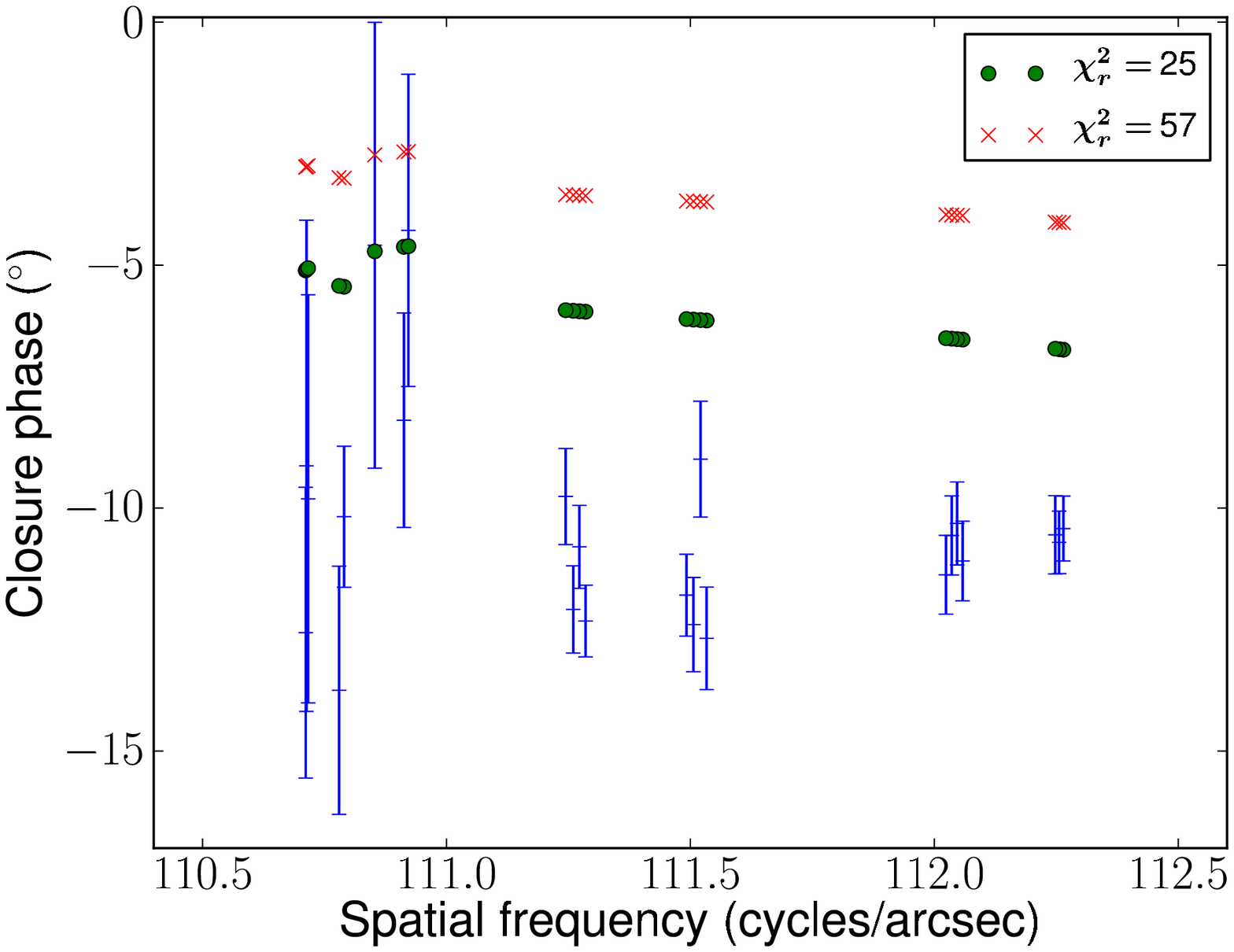} 
   \hspace{1cm}
   \includegraphics[width=8cm,height=7cm]{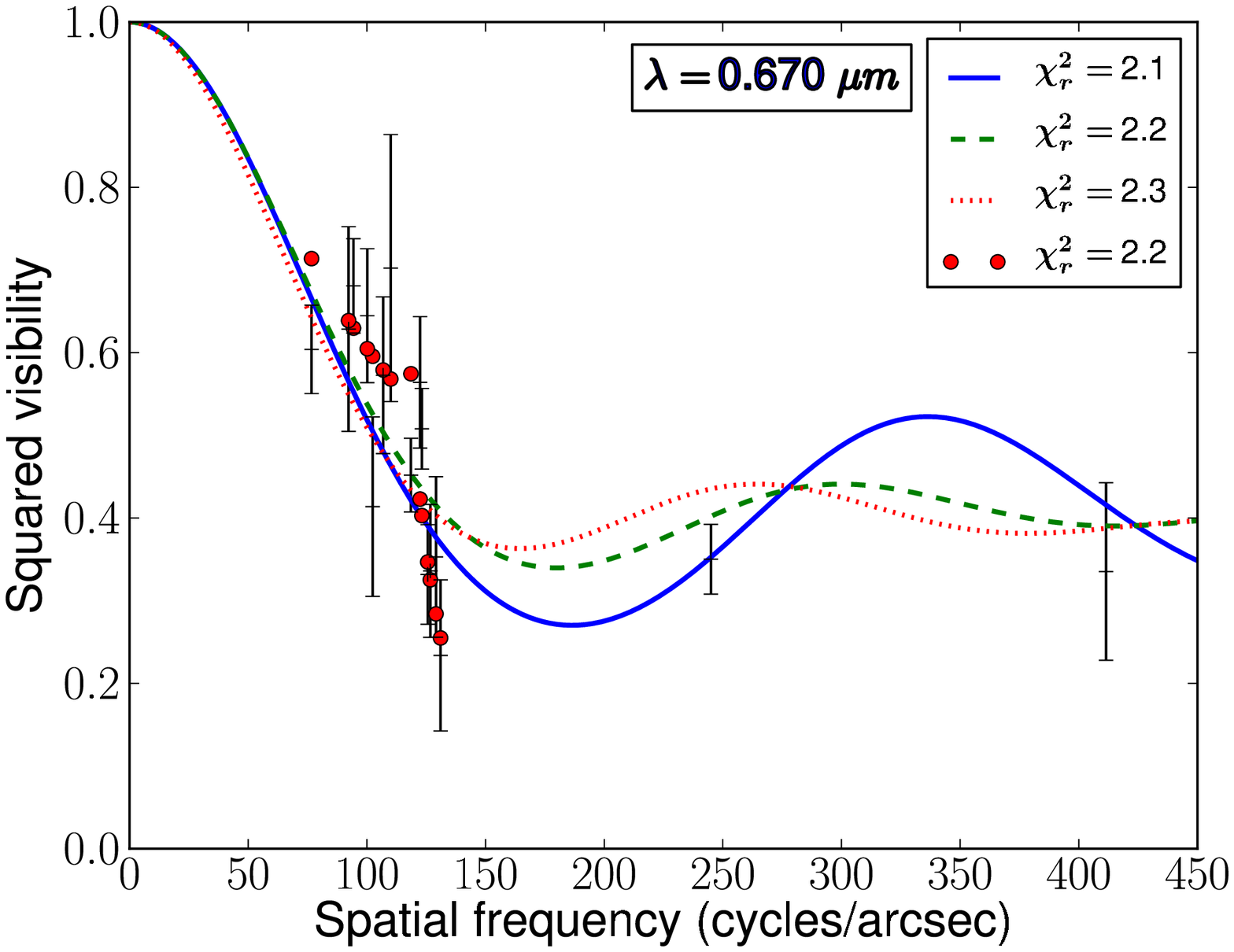}
     \caption{Left panel: the 2003 IOTA closure phases on which the astrometric fit of Sect.~\ref{subsection:near-IRimprovement} was based,
     together with two models, as a function of the maximal spatial frequency in the baseline triangle. The green circles are 
     for a model including only the post-AGB star offset to a position (x,y) = (-0.4,-0.4)~mas. The red crosses illustrate the effect 
     of adding the secondary at a position (+0.4,+0.4)~mas and contributing 10\% of the total flux. Right panel: 
     squared visibilities at 673~nm with some models overplotted. The blue, green, and red lines correspond to star+ring models with ring diameters and widths of 
     (5.5,1.0), (3.0,3.0), and (1.0,4.5), respectively. The red filled circles correspond to the best binary model deduced in 
     Sect.~\ref{subsection:opticalbinary} and online Fig.~\ref{figure:chi2map-opticalbinary}. The other wavelengths are shown online in Fig.~\ref{figure:allOptVis}. 
     The legend lists the $\chi^2_r$ value, which is the same for all four models.}
     \label{figure:closures+optical}
\end{figure*} 

\citet{2002ApJMonnier} argued that fitting near-IR visibilities of protoplanetary disks with a uniform ring model is well
justified: in the standard model of a passive disk, the near-IR emission is dominated by the hottest dust at the sublimation radius. 
This simple model works well if the inclination is small and if the region between the central star and the inner rim 
is optically thin. The first condition is certainly fulfilled for 89\,Her (with i$\sim$10-15$^\circ$), so we 
use such a pole-on star+ring model. In Sects.~\ref{subsection:opticalbinary} and~\ref{subsection:near-IRimprovement}, we discuss 
the properties of the central binary and how it affects our observations in more detail; 
here we assume a single star at the center of the ring. Our model parameters 
are the angular diameter of the post-AGB star $\theta_\star$, the inner diameter of the ring $\theta_{\mathrm{in}}$, the width of the ring 
$\Delta (\theta_{\mathrm{in}} / 2)$, and the flux ratio of the two components 
$f_\star=F_\star/F_{\mathrm{tot}}$ and $1-f_\star=F_{\mathrm{ring}}/F_{\mathrm{tot}}$. 
The visibility is computed as $V = f_\star V_{\star} + (1-f_\star) V_{\mathrm{ring}}$.

Table~\ref{table:LITpro} lists the parameter space that was examined. 
The good uv coverage allows the size and flux contribution of the uniform ring to be constraine, independently of any assumptions or spectral fitting. 
A first estimate of the flux ratio was determined 
from the height of the visibility plateau at the highest spatial frequencies, where V$_{\mathrm{ring}} \sim 0$ and V$_{\star} \sim 1$. The K band spatial
resolution is insufficient to be sensitive to $\theta_\star$. % within the given ranges.
Stellar diameters $>$0.6~mas only fit the H band data when combined with large stellar flux contributions and specific 
ring sizes (so that its higher lobes compensate the visibility drop due to the large $\theta_\star$) 
and are assumed to be local minima. The H band data are most compatible with $\theta_\star = 0.45\pm0.15$~mas, 
which is also found from SED fitting in Sect.~\ref{section:stellarsed}, and we fix this value in what follows. 
The global $\chi^2_r$ minimum was found through a Levenberg-Marquardt optimization, starting 
from the best minimum found in the $\chi^2_r$ maps of the ring parameters for a range of flux ratios.

\begin{figure*}
\centering
   \includegraphics[width=18cm]{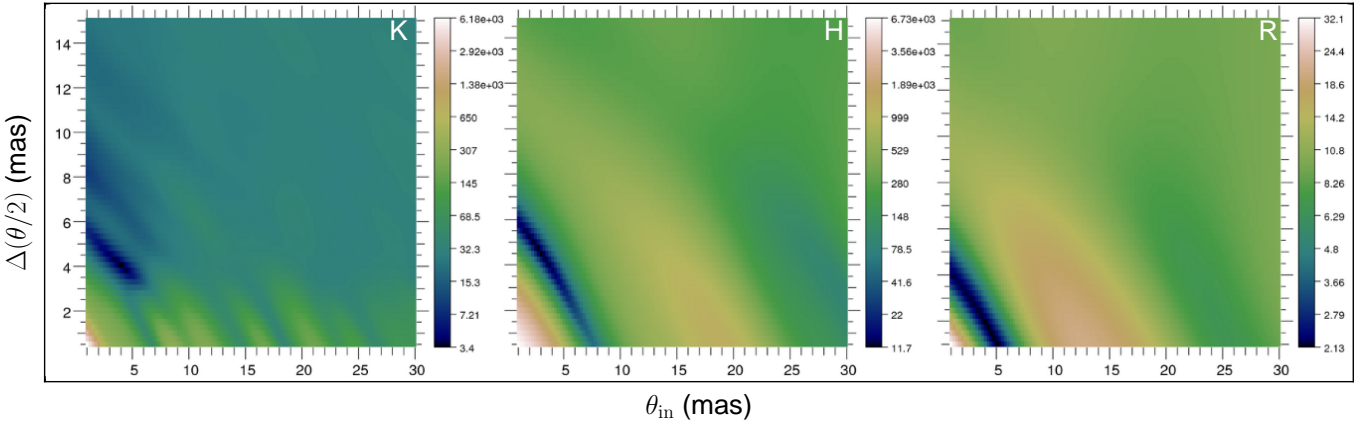} %RingdiskFstar055thetastar044chi2map_AMBERCHARAPTI.eps}
     \caption{$\chi_r^2$ maps of the ring's geometric parameters for the star+ring model that was fitted to the three different visibility data sets. From left to right: 
     K, H and R-band.}
     \label{figure:chi2maps1}
\end{figure*}

The best-fit values are listed in Table~\ref{table:LITpro}. 
The $\chi^2_r$ maps for the best-fit flux ratios are shown in Fig.~\ref{figure:chi2maps1} and the best-fit model visibility curves 
in Fig.~\ref{figure:Hsinglering} and~\ref{figure:Ksinglering} (available online). Their left panels also contain the 
AMBER 2007 data in red, to which a fit was made in the H band by fixing $\theta_{\mathrm{in}}$ to 1~mas and leaving the 
flux ratio and ring width free. This gives a slightly increased stellar flux contribution of 60\% (vs. 55\% in the general best-fit model) 
and a ring width of 4~mas, similar to that in the optical (see Sect.~\ref{subsection:opticalring}).

The simple model fits the observations well, as is evidenced by the low $\chi^2_r$ values. The best-fit ring is similar
in the H and K bands. Particularly in H, but also in K band, a small degeneracy still exists between the ring's inner diameter and its width,
with a slight preference for smaller diameters in combination with larger widths. 
This means there is flux coming from radii close to the binary orbit,
%\footnote{$\theta_{\mathrm{in}}=4$~mas and $\Delta (\theta_{\mathrm{in}} / 2)=4$~mas, gives ring radii from 2 to 6~mas.}
which is in contrast to our original assumption that 
the emission is coming from dust in a well-localized inner rim. A wide uniform ring works well but is not perfect.

\onlfig{6}{
\begin{figure}
\centering
   \includegraphics[width=8cm,height=12cm]{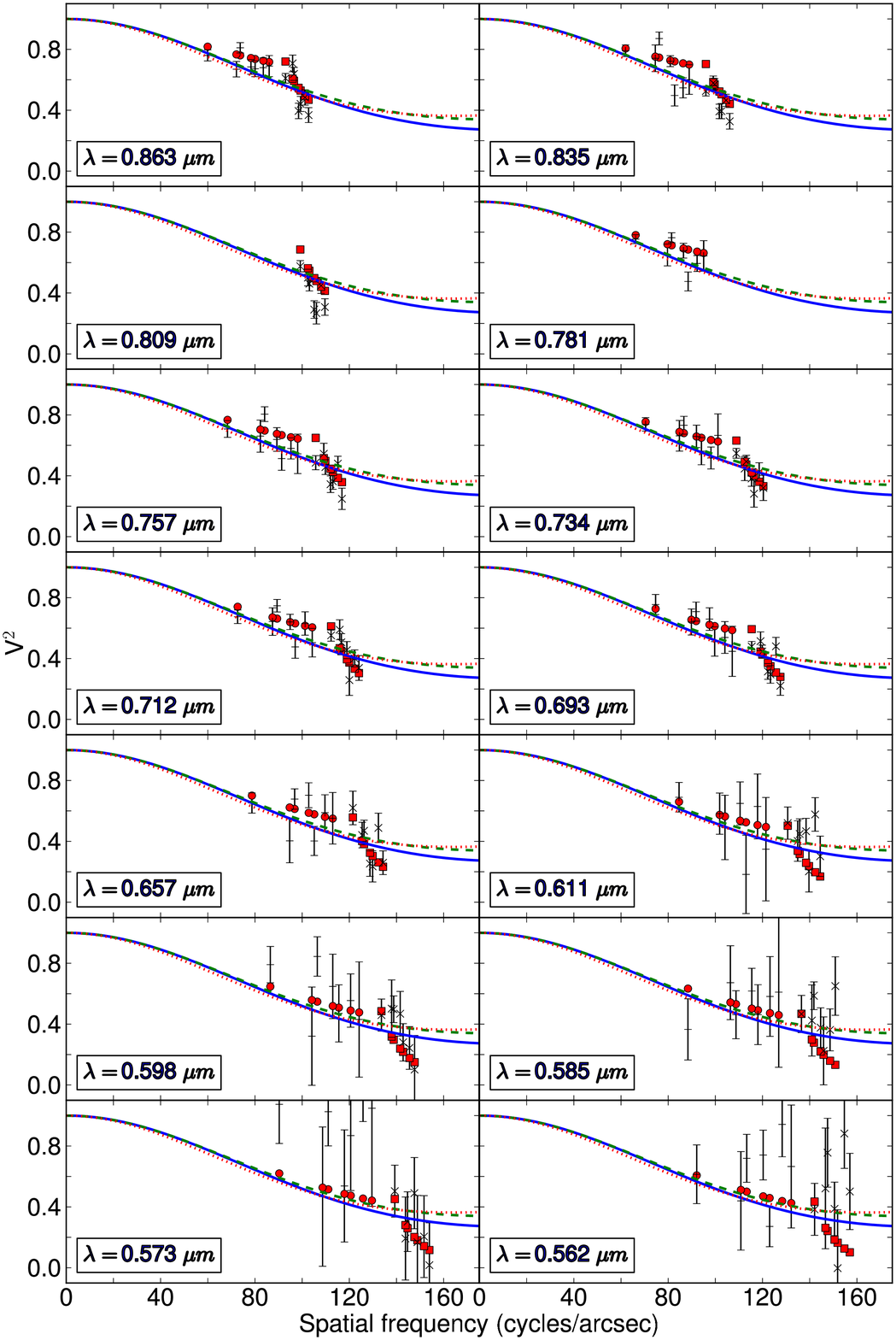} %{allOptVis.eps}
   %\hspace{1cm}
     \caption{All NPOI visibilities, except the 673~nm channel which is shown in Fig.~\ref{figure:closures+optical}. A different 
     wavelength is shown in each panel, as indicated. The symbols correspond to different physical baselines, plusses for AC0-AE0 and crosses for AE0-AW0.
     The blue, green and red lines are the same star+ring models as in Fig.~\ref{figure:closures+optical}, and have ring diameters and widths of 
     (5.5,1.0), (3.0,3.0) and (1.0,4.5) respectively. The red filled circles correspond to the best binary model deduced in 
     Sect.~\ref{subsection:opticalbinary} and online Fig.~\ref{figure:chi2map-opticalbinary}.}
     \label{figure:allOptVis}
\end{figure} 
}

\onlfig{7}{
\begin{figure}
\centering
   %\includegraphics[width=8cm,height=7cm]{89Her_geometric_673nmV2.eps} %{Diskdisk0673Fstar068Thetastar045Thetadisk05bestsepv2.eps} %Disk-disk_0.673_Fstar0.68-Thetastar0.45-Thetadisk0.5_bestsep_v2.png
   %\hspace{1cm}
   \includegraphics[width=8cm,height=7cm]{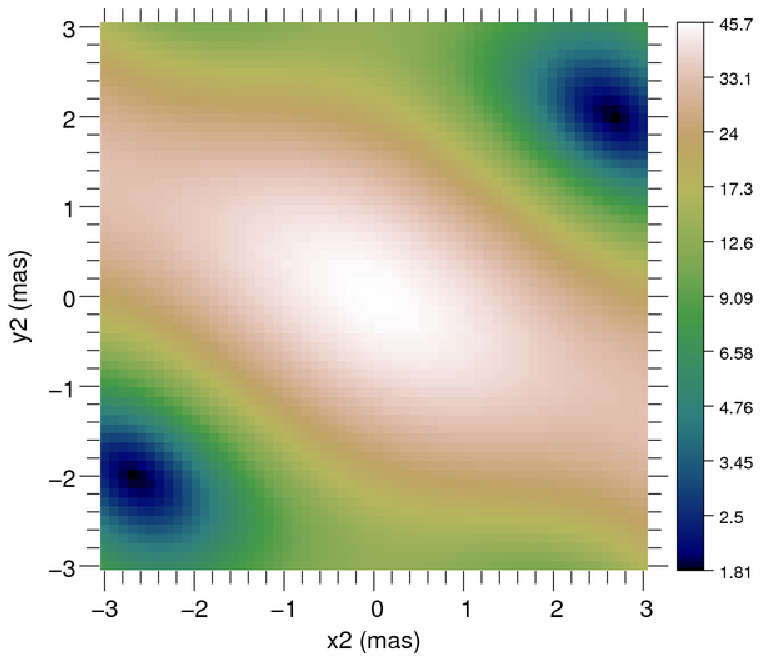}  %{Diskdisk056206730859Thetastar045Thetadisk05bestsep.eps} %Disk-disk_0.562-0.673-0.859_Thetastar0.45-Thetadisk0.5_bestsep.png
    \caption{The $\chi^2_r$ map of the relative position of the secondary component with respect to the primary at the center, 
     if a binary model is applied to fit the optical NPOI visibilities.}
     \label{figure:chi2map-opticalbinary}
\end{figure} 
}

\subsection{The central binary in the optical?}\label{subsection:opticalbinary}
As shown in Fig.~\ref{figure:closures+optical}, our measured optical continuum visibilities are lower than expected for a uniform 
disk (UD) of $\theta_\star=0.45$~mas and do not have the spatial frequency dependence of such a simple geometry.
It is unlikely that we resolve the binary as the companion is probably
still on the main sequence and hence too faint.  
However, accretion disks, already inferred from spectral monitoring by several
authors for similar objects \citep[see, e.g.,][]{2009ApJWitt,2012AAGorlova}, are typically hot and can
contribute significantly at optical wavelengths, yielding a binary-like signature in the observations.   
Because the VEGA and NPOI data were obtained at different orbital phases of the binary system, we analyzed the
most extensive snapshot, i.e., the NPOI data, first. Already from this analysis, the binary model can be ruled out, as detailed below.

% \begin{figure*}
% \centering
%    \includegraphics[width=8cm,height=7cm]{Ringdisk_Fstar056_thetastar0435chi2map_IOTAAMBERCHARAPTI.eps} %RingdiskFstar055thetastar044chi2map_AMBERCHARAPTI.eps}
%    \hspace{1cm}
%    \includegraphics[width=8cm,height=7cm]{Ringdisk_Fstar028-thetastar0435chi2map_AMBERCHARAPTI.eps} %RingdiskFstar025thetastar044chi2map_AMBERCHARAPTI.eps} 
%      \caption{$\chi_r^2$ maps of the ring's geometric parameters for the star+ring model that was fitted to the near-IR visibilities. Left: H band, right: K band.}
%      \label{figure:chi2maps1}
% \end{figure*} 

Since the flux ratio can be wavelength dependent, we selected a subset of three wavelengths that cover the full spectral range, 
i.e., at 562, 673, and 859~nm. Given the lowest squared visibility of $\sim$0.2 (Fig.~\ref{figure:closures+optical}), the flux ratio 
$F_{\star}/F_{\mathrm{sec}}$ should be well within the range $\{0.2,5.0\}$. Starting from equal fluxes and assuming diameters 
of 0.45 mas for the post-AGB star and 0.5~mas for the accretion disk,
we search for the best relative position of both components. Then, the
flux ratios and position are iteratively adjusted to find the global minimum. The data do not constrain the flux ratios well, 
but the exact values do not influence the minimal separation needed to fit the data. Thanks to the good baseline position angle coverage, 
the positional minimum is rather well defined, except for the 180$^\circ$ ambiguity that can only be resolved with closure phases. 
The minimal separation is 2~mas (for flux ratios of 5.0, 2.2, and 1.4 at 562, 673, and 859~nm, respectively), and 
this value does not depend on the exact post-AGB and accretion disk diameters. The squared visibilities of the best-fit binary model 
are displayed in Figs.~\ref{figure:closures+optical} and~\ref{figure:allOptVis}, and the $\chi^2_r$ map of the secondary's relative 
position is shown in Fig.~\ref{figure:chi2map-opticalbinary}.

At a distance of 1.5$\pm$0.5~kpc, the required angular separation converts into a minimal physical separation of $\rho=2$\,AU at the NPOI epoch. 
The smallest semi-major axis that can accomodate such a projected separation is for an orbit in which the time of the NPOI measurement would correspond to apastron passage 
and with $\omega=0^\circ$. In that limiting case, also assuming the maximal eccentricity of 0.25 (see Table~\ref{table:basicpars}), $a=\rho/(1+e)=1.6$\,AU. 
For 89 Her's period of 288 days, this implies a minimal total system mass of 6.6~M$_\odot$. If the primary is a post-AGB star of $\sim$0.6\,M$_\odot$, 
the inclination is $3.2^\circ$, as derived from the spectroscopic mass function.

However, such a high total mass cannot be reconciled with a post-AGB status for the primary because it would 
require an extremely efficient mass transfer as well as an original primary mass that is incompatible with the system's high Galactic latitude and the absence 
of a detected ejection velocity. % The latter also argues against a misclassification of the primary as a post-AGB star.}
In addition, there is no signature of the companion's velocity in the optical spectra 
\citep{1993AAWaters,1987IAUSClimenhaga}, which is incompatible with
the required (from the flux ratio) high luminosity for this putative
high-mass companion or its accretion disk. At the minimal distance of 1~kpc, the secondary's luminosity would need to be $\sim$2100\,L$_\odot$ 
(see Sect.~\ref{section:stellarsed}). For a 6\,M$_\odot$ main sequence star of radius 3\,R$_\odot$, this implies an accretion luminosity of $\sim$1350\,L$_\odot$
and hence an accretion rate of $LR/GM=2 \times 10^{-5}$~M$_\odot$~yr$^{-1}$. This rate would be too high to be compatible with the mass loss rate derived 
from the P Cygni type H$\alpha$ profile of $\sim$10$^{-8}$~M$_\odot$~yr$^{-1}$.

 \begin{table}
\caption{Parameter space for a star+ring model searched with LITpro, and the final best-fit values.}             % title of Table
\label{table:LITpro}      % is used to refer this table in the text
\centering                          % used for centering table
\begin{tabular}{c c c c c c}        % centered columns (4 columns)
\hline\hline                 % inserts double horizontal lines
Parameter & Range & Best H & Best K & Best optical\\    % table heading 
\hline                        % inserts single horizontal line
   F$_\star$/F$_{\mathrm{tot}}$ & (0.0,1.0) & 0.56 $\pm$ 0.05 & 0.28 $\pm$ 0.05 & 0.65 $\pm$ 0.07 \\      % inserting body of the table
   F$_{\mathrm{ring}}$/F$_{\mathrm{tot}}$ & (0.0,1.0) & 0.44 $\pm$ 0.05 & 0.72 $\pm$ 0.05 & 0.35 $\pm$ 0.07 \\
   $\theta_\star$\tablefootmark{a} (mas) & (0.2,0.7) & 0.45$\pm$0.15 & 0.435\tablefootmark{b} & 0.435\tablefootmark{b}\\
   $\theta_{\mathrm{in}}$\tablefootmark{a,c} & (1.0,30.0) & 2.5 $\pm$ 1.0 & 4.1 $\pm$ 0.4 & 3 $\pm$ 2 \\
   $\Delta (\theta / 2)$\tablefootmark{a,c} & (1.0,15.0) & 4.5 $\pm$ 1.0 & 4.0 $\pm$ 0.2 & 3 $\pm$ 2 \\
   $\chi^2_r \pm \sigma$\tablefootmark{d} & - & 11.8 $\pm$ 0.1 & 3.8 $\pm$ 0.3 & 2.1 $\pm$ 0.1 \\
\hline                                   %inserts single line
\end{tabular}
\tablefoot{\tablefoottext{a}{$\theta$ denotes diameter (mas), so $\Delta (\theta / 2)$ is the width of the ring as the difference between its outer and 
inner radius.} \tablefoottext{b}{Data impose no constraint, so value fixed from SED fit.} \tablefoottext{c}{A strong correlation exists between these parameters, so the 
given errors reflect the range of good $\chi^2_r$. We refer to the $\chi^2_r$ maps for a better estimate of the confidence interval.} \tablefoottext{d}{This is 
the final $\chi^2_r$ obtained by fitting all parameters, as opposed to the $\chi^2_r$ maps where only the shown parameters are left free. So the number 
of degrees of freedom is slightly different in both cases.}}
\end{table}

Furthermore, although the inclination of \citet{2007AABujarrabal} 
is probably not precise to a few degrees, a value of 3$^\circ$ is hard
to reconcile with the asymmetry observed in their CO maps. 
Finally, the near-IR visibilities do not allow for the presence of
a low-contrast second stellar-like component (see Sect.~\ref{subsection:near-IRimprovement}), and no
star (or accretion disk at a realistic temperature) is blue enough to be optically bright while being indetectable in the near-IR.
We thus conclude that the spatial scale and flux ratio of the central binary do not agree with what is needed to fit our optical interferometry data.

\subsection{A ring in the optical}\label{subsection:opticalring}
The small inner diameter of the best-fit near-IR ring and the similar scale found with the binary model motivated us to 
try a ring model to explain our optical data as well. The geometric star+ring model of Sect.~\ref{subsection:nearIRring} was fit 
to all the VEGA and NPOI data. Ring diameters and widths up to 50~mas were tried to allow for a strong scattered light contribution from the outer disk. 

The longer baseline VEGA observations are essential to resolve some of the degeneracies between the model parameters, especially to fix the flux ratio more precisely. 
The fitting strategy was the same as in Sect.~\ref{subsection:nearIRring}. We adopted a single flux ratio at all wavelengths (see Sect.~\ref{subsection:HERMES}). 
This resulted in a good $\chi^2_r$, while a wavelength-dependent flux ratio does not improve the fit.

Three model visibility curves and the $\chi^2_r$ map of the ring diameter versus its width are 
depicted in Figs.~\ref{figure:closures+optical} and~\ref{figure:chi2maps1}, respectively.
A uniform ring fits the optical observations very well. The size of the emitting area is rather small with an outer radius 
of only 3.5 to 5.5~mas. This agrees well with the fit to the AMBER 2007 data.
The inner diameter is not constrained. By changing the flux ratio slightly, the same $\chi^2_r$ can be obtained both 
for an infinitely thin ring at the outer edge (a circle) and for a ring starting at the binary orbit (a disk), and part of the emission 
may even originate from within it. Independent of the inner diameter, the outer radius of the ring is smaller than in the near-IR. 
%It is possible that part of the emission comes from within the binary orbit. 

\subsection{Improving the near-IR fit}\label{subsection:near-IRimprovement}
A circumcompanion accretion disk is not causing the low optical visibilities but could still affect our near-IR data, providing it has a low 
luminosity and temperature. Although we have good $\chi^2_r$ values, the best-fit is still not perfect and results in a rather small inner diameter. 
A closer look at the near-IR data offers a few arguments against a signature of the secondary in it.  

A strong constraint on the secondary's near-IR flux contribution comes from the lack of visibility variations 
over the orbital cycle on the two PTI baselines. As the binary moves in its orbit, the projected separation onto the 
fixed baseline changes, thereby modulating the visibility. The visibility scatter then gives a lower limit on the contrast of the binary.
The strongest constraint is derived in the H band because the circumbinary disk dominates the K band.
With a continuous phase coverage over a third of the orbital cycle in 2001, we estimate a contrast $>$5 in the H band.

The strongest contribution to the residuals of the best-fit star+ring model comes from the shortest and longest AMBER baseline in K and from the
two shortest IOTA baselines in the H band. A simple way to improve the general fit would be to add a small ($\sim$5\%) flux component 
that is fully resolved already at our shortest spatial frequencies. This relaxes the visibility slope in the first lobe and fits 
with the interpretation of the AMBER 2007 data as being erroneous even in the K band. %Another way could be to make the emission profile of the ring smoother, more Gaussian-like. 
We note that a proper radiative transfer model, which takes scattering fully into account, typically has such a feature 
\citep{2008ApJPinte}. This is because material at radii beyond the inner rim scatters part of the inner rim thermal emission into the line of sight. We note that
although part of this flux might fall outside the coherent FOV (=$\lambda^2/(B \Delta \lambda)$ $\sim$20~mas) at the longest CLIMB baselines, this does not 
influence any of our results as the dust disk is already fully resolved at these baselines. Our estimation of the flux ratio is thus robust.

\onlfig{10}{
\begin{figure}
\centering
   \includegraphics[width=7cm,height=6cm]{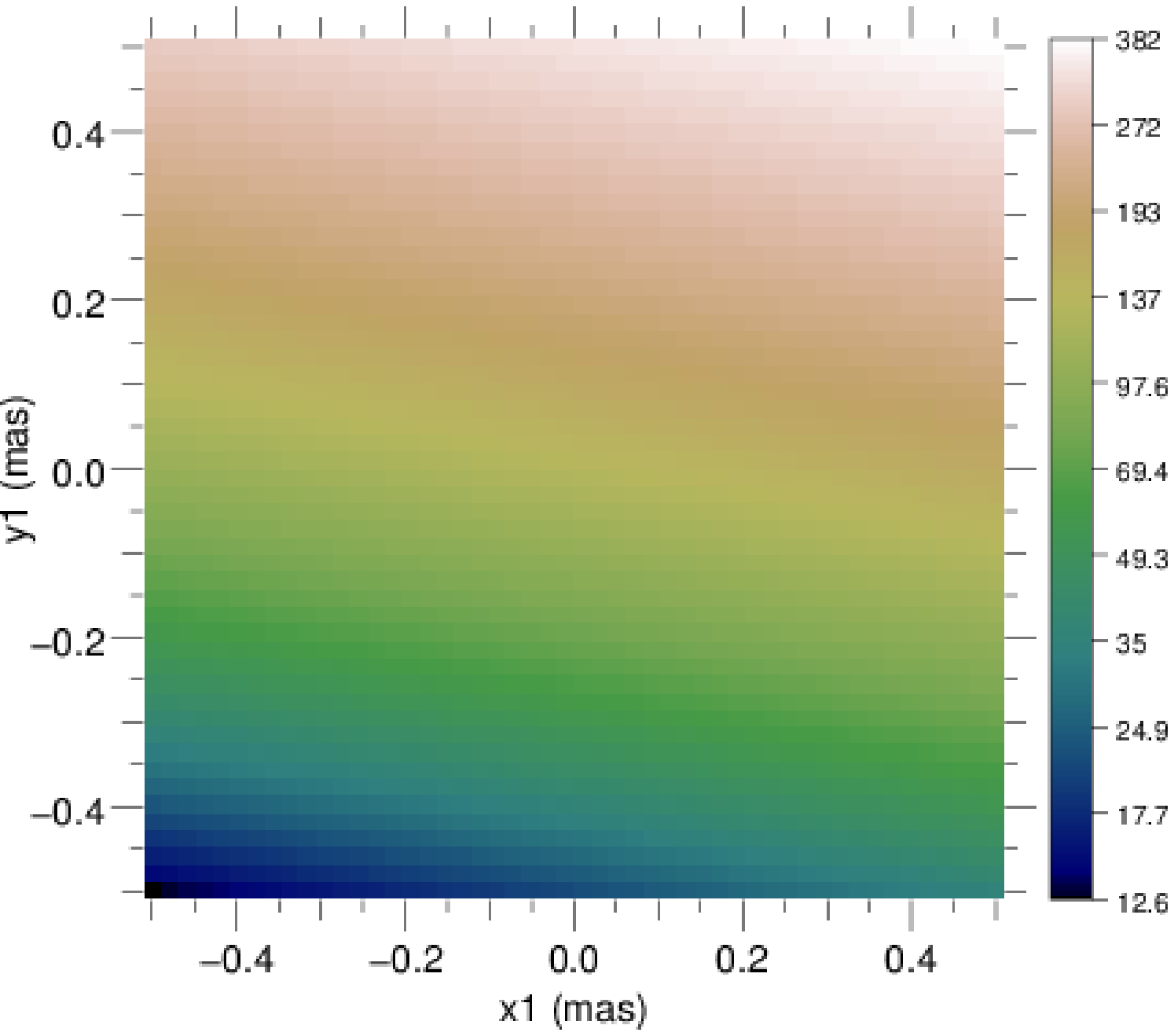}
     \caption{The $\chi^2_r$ map of the position x versus y of the post-AGB star with respect to the phase reference defined by the ring 
     (inner diameter 3~mas and width 4.6~mas, see Sect.~\ref{subsection:near-IRimprovement}).}
     \label{figure:chi2closure}
\end{figure} 
}

Second, we checked the compatibility of the closure phases with the expected post-AGB offset from 
the phase reference provided by the ring. No constraints could be derived from the AMBER and CLIMB closures 
due to their large errors, so we present our analysis of the more numerous and precise IOTA data. 
The expected range of displacements from the center of mass is 0.2-0.5~mas, depending on the distance, spectroscopic orbit, 
and inclination. Fig.~\ref{figure:chi2closure} (available online) shows the $\chi^2_r$ map of the IOTA closure phases as a function of the post-AGB star's 
position with respect to the best-fit H band ring. Even at the maximal allowed offset, there is a significant 
difference between the measurements and the model ($\chi^2_r = 12.6$).
The measurements are depicted in the left panel of Fig.~\ref{figure:closures+optical}, along with two models. 
The green circles correspond to a model in which the post-AGB star is positioned at (-0.4,-0.4)~mas, which is more realistic 
than the boundary values preferred by the fit.
For the red crosses, the secondary is added at the opposite position as well and contributes 10\% of the total flux (corresponding to a flux ratio of $\sim$5). 
The red model shows that if the secondary is present at the limiting flux contribution determined above, 
even at a modest separation, the discrepancy between the model and measured closure phases becomes significantly worse.

There are several possibilities to explain the remaining discrepancy. First, the expected displacement could be underestimated, which could imply an 
overestimated distance. Second, at the given spatial frequencies, the measured closure phases will be 
strongly influenced by the exact geometry of the circumbinary disk and any deviation from point symmetry in it. Here we approximate the disk  
with a uniform ring, whose size and width influence the closure phases directly through the cross-terms 
in the triple product. Given the small inclination, the skewness of the disk is 
insufficient to induce a strong closure phase signal if it is inherently point-symmetric. An intrinsic asymmetric 
illumination of the disk might be required to explain the measured closure phases. This would not be hard to imagine, given the close proximity of the 
best-fit ring to the central object. To properly take these effects into account at the precision level of the IOTA measurements, 
a detailed radiative transfer model is needed, so we did not try to improve the $\chi^2_r$ within the current modeling framework.

\subsection{Summarizing remarks}
We summarize our interferometric analysis as follows:
\begin{enumerate}
 \item Our H, K, and optical visibilities, which were collected with different instruments and span a decade in time, are fully 
 consistent with a simple geometric star+ring model. 
 \item The central post-AGB source only contributes 28$\pm$5\%, 56$\pm$5\%, and 65$\pm$7\% of the total received flux at 2.2, 1.65, and 0.67~$\mu$m respectively.
 \item Our near-IR data exclude a geometrically thin ring. Part of the emission might stem from close to the central binary.
 \item The optical ring can be either geometrically thin or extended, but its outer diameter is smaller than the near-IR one.
 \item A low-contrast binary model can fit equally well the optical visibilities but results in an unrealistic binary configuration.
 \item Our largest spatial frequencies in H and K, the time series of PTI H visibilities, and the IOTA closure phases rule out the presence of a 
 companion with a contrast $\leq$5.
 \item It is unclear whether our closure phase discrepancy is caused by an inconsistency in the standard orbital parameters, the radial intensity distribution
 of our uniform ring model, or an azimuthally asymmetric illumination of the inner disk rim. 
 %More precise, phase-resolved, closures might allow to determine the orbital motion of the primary with respect to the center-of-mass of the system.
 \item The 2007 AMBER data deviate from all the rest, especially in H and at the shortest baselines. The latter makes an astrophysical 
 interpretation problematic. %We simply remark the similarity between its best-fit ring size and that obtained in the optical.
 \end{enumerate}

%______________________________________________________________

\onlfig{11}{
\begin{figure*}
\centering
  \includegraphics[width=8cm,height=7cm]{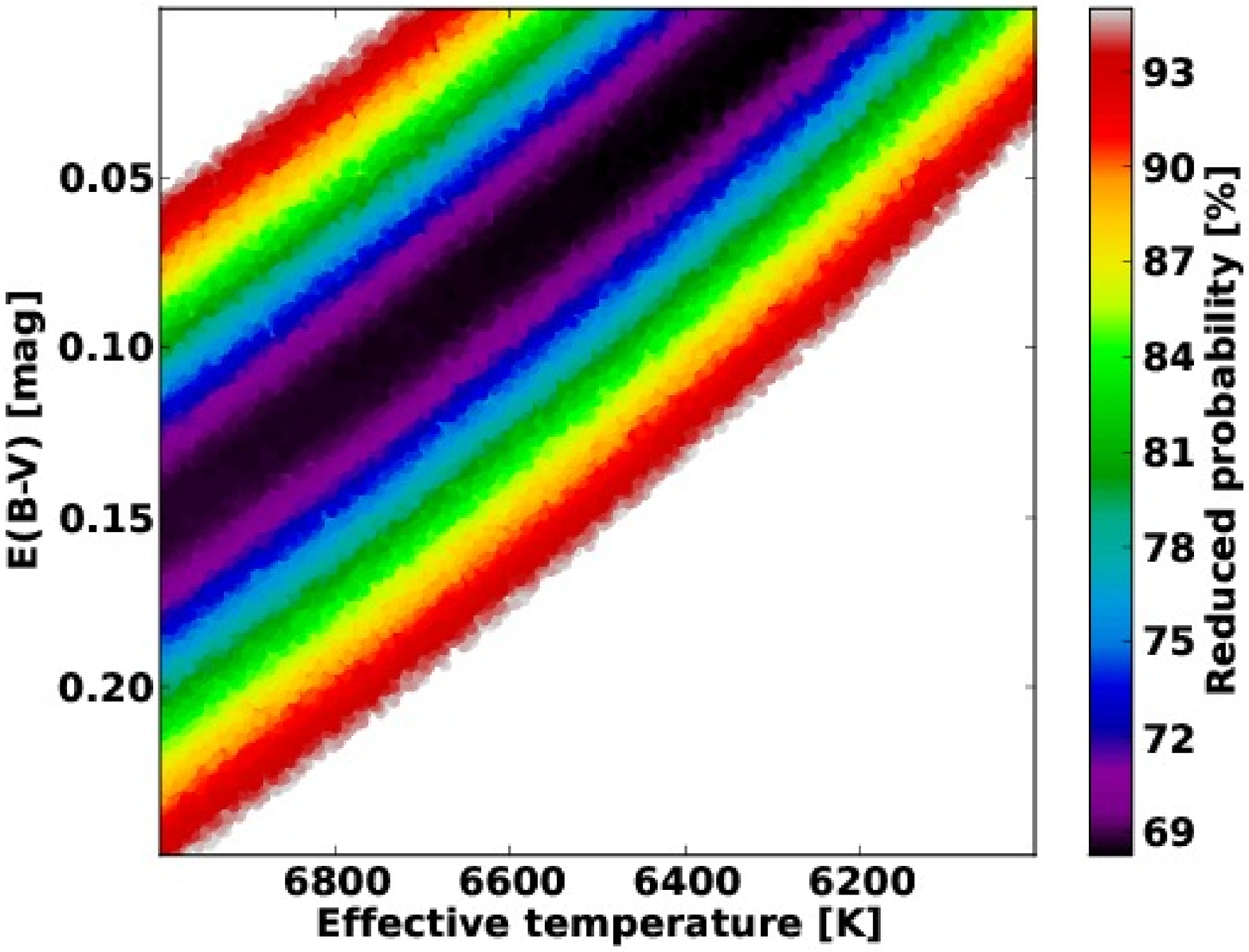}  %{HD163506_igridebv.eps}
  \hspace{1cm}
  \includegraphics[width=8cm,height=7cm]{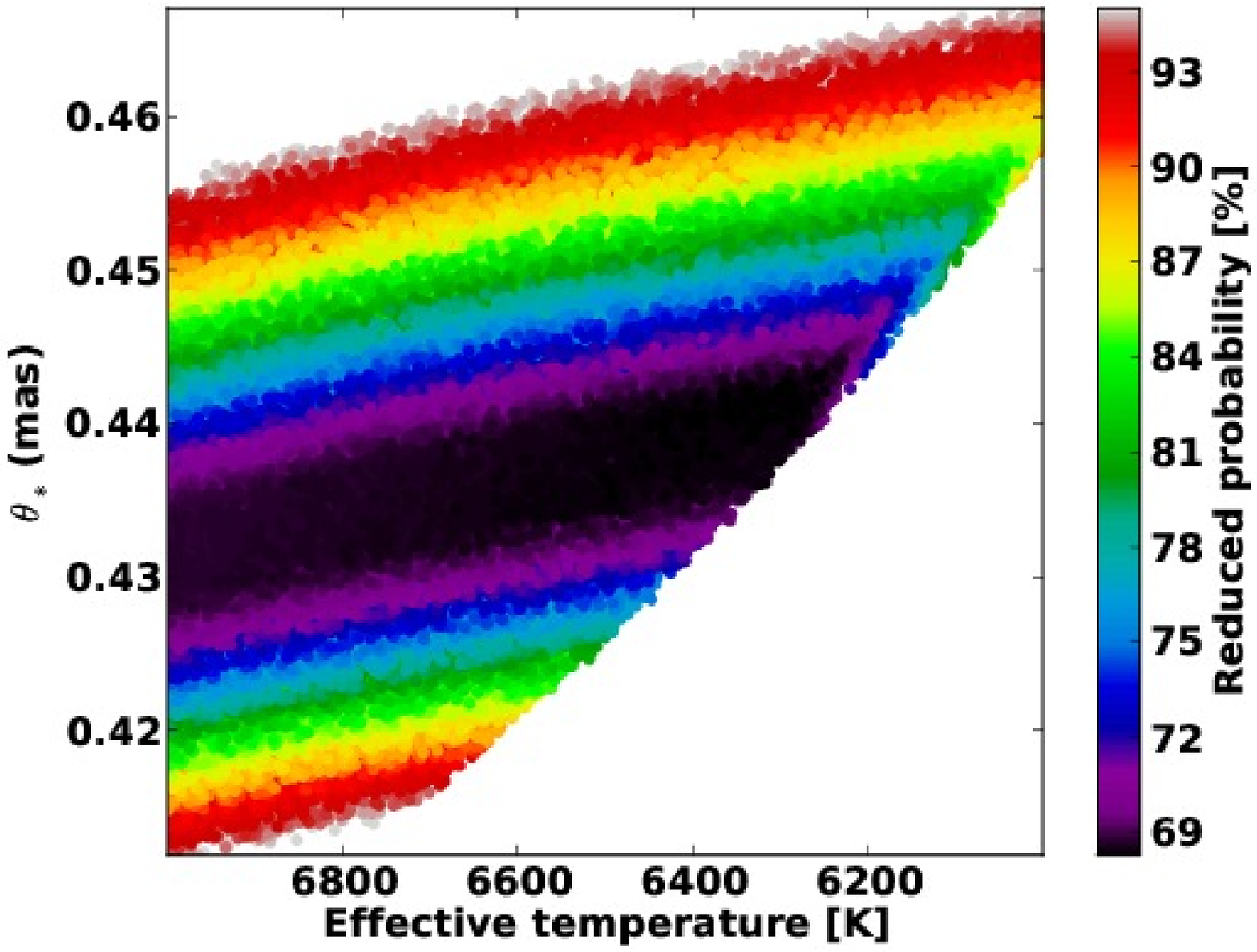}   %{HD163506_igridthetastar.eps}
    \caption{Left: CI of T$_{\mathrm{eff}}$ versus E(B-V), right: CI of T$_{\mathrm{eff}}$ versus angular diameter $\theta_{\star}$ (mas), for 
    the stellar SED fit.}
    \label{figure:chi2map-SED}
\end{figure*} 
}

\begin{figure*}
\centering
   \includegraphics[width=8cm,height=7cm]{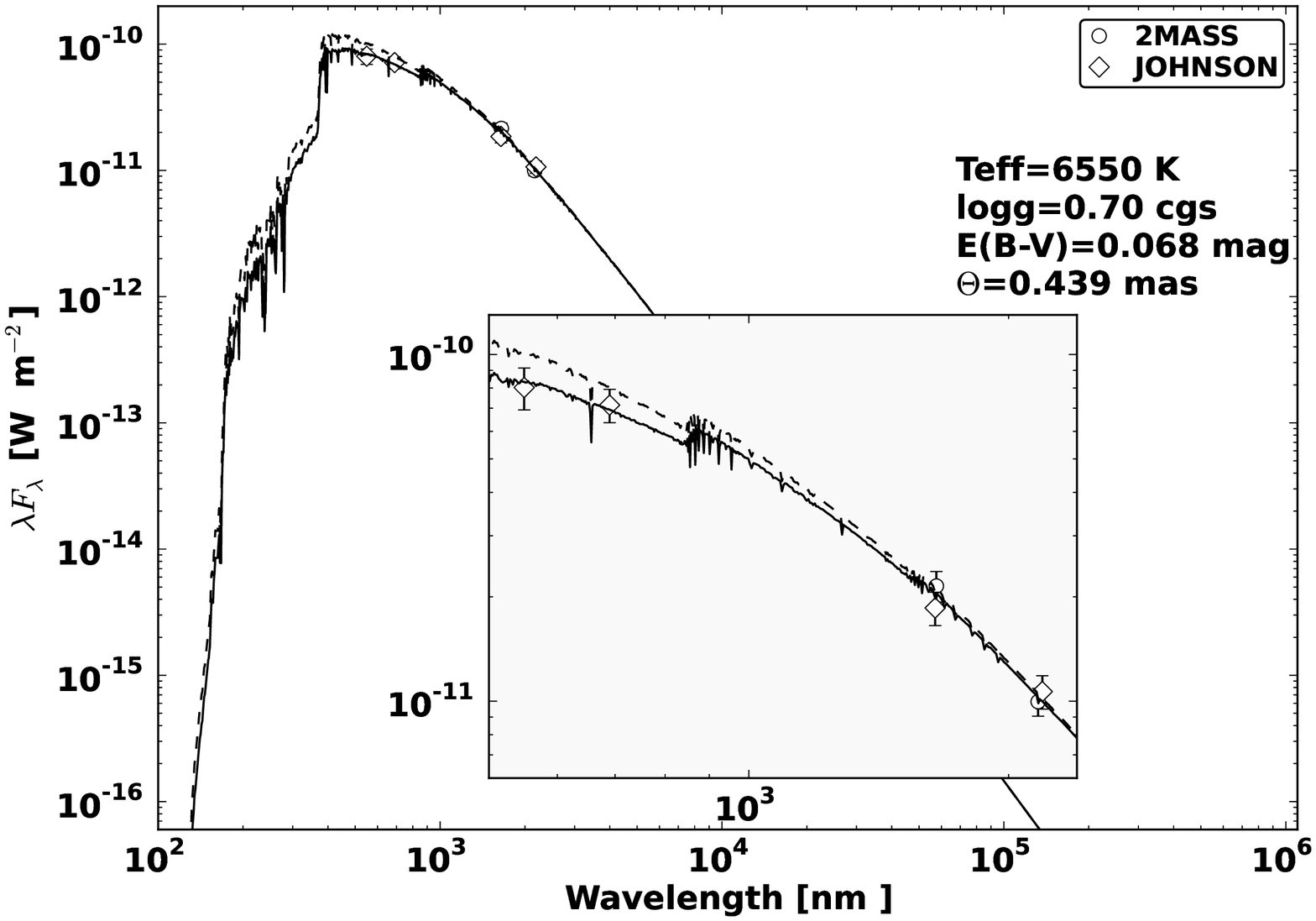}  
   \hspace{1cm}
   \includegraphics[width=8cm,height=7cm]{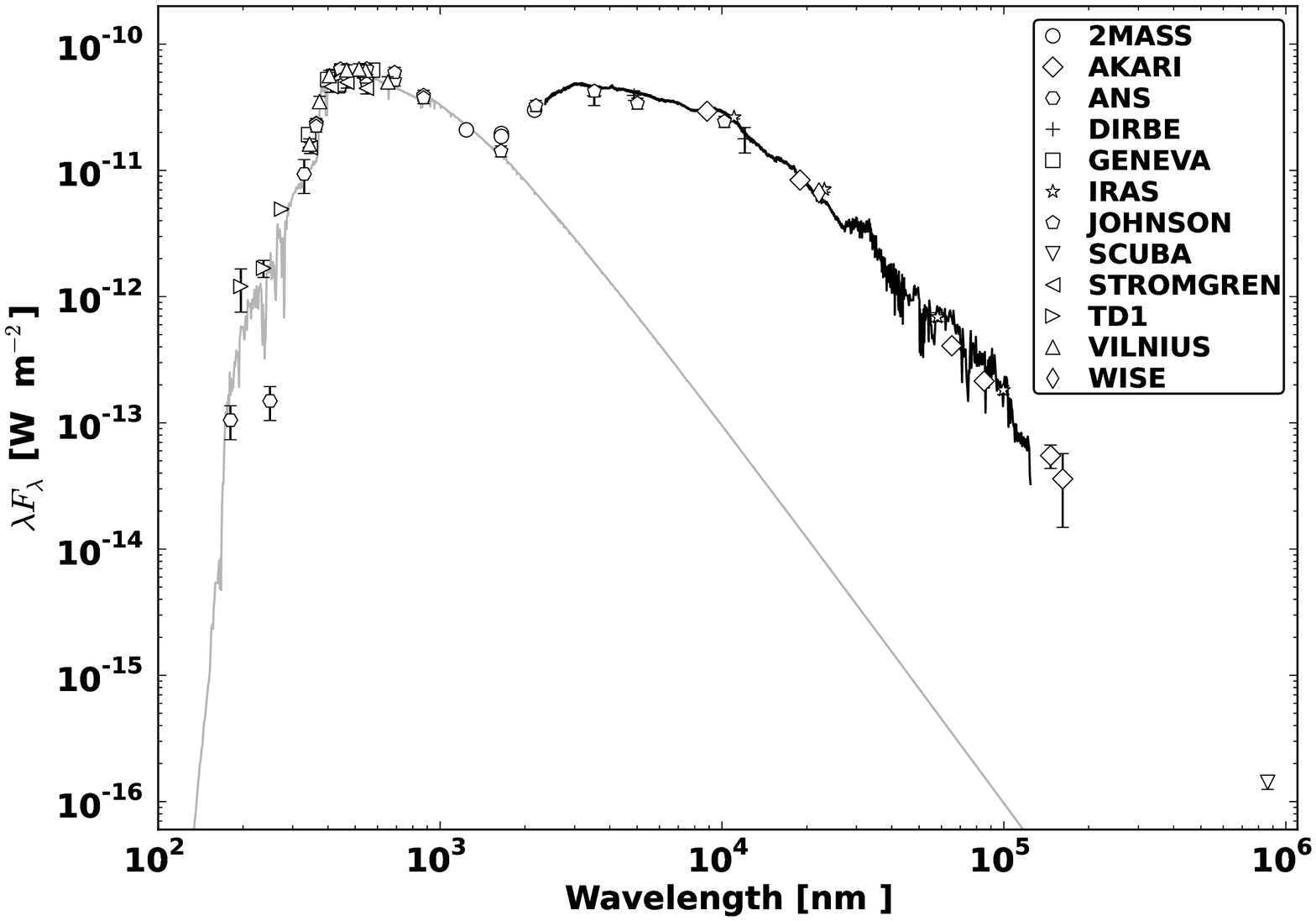}
     \caption{Left: the observed (i.e., reddened, in full) and original (i.e., dereddened, in dashed) stellar SED model plotted over the stellar fluxes. 
     The inset shows a zoom-in on the actual measurements.
     Right: the circumstellar SED (not corrected for reddening), obtained after subtracting the measured total fluxes with the reddened stellar Kurucz model, 
     overplotted with the same reddened Kurucz model (in gray) but with the angular diameter arbitrarily scaled to these circumstellar fluxes. Both panels
     have the same scales on the x and y axes to illustrate the large amount of circumstellar flux.}
     \label{figure:SED}
\end{figure*} 

\section{The SED}\label{section:stellarsed}
\subsection{The stellar SED}
In our search for the origin of the resolved optical flux, we first re-evaluated the stellar parameters. ``Stellar
fluxes'' are obtained by combining the interferometric flux ratio (Table~\ref{table:LITpro}) with the total photometric fluxes. 
Four photometric bands (V, R, H, and K) were used to refit the SED. This is insufficient to independently constrain all model parameters, i.e., 
the effective temperature T$_{\mathrm{eff}}$, surface gravity $\log$~g, metallicity z, interstellar reddening E(B-V), and angular diameter $\theta_\star$. So we 
restricted some parameters by taking spectroscopic constraints into account, as listed in Table~\ref{table:basicpars}. We fix the metallicity to the subsolar 
value of $[\mathrm{Fe}/\mathrm{H}]=-0.5$ and bound $\log$~g to \{0.3,0.8\}~dex, T$_{\mathrm{eff}}$ to \{6000,7000\}~K, and E(B-V)$<$0.25~mag. 

Since the optical interferometric data correspond to the JOHNSON R band, while the absolute magnitude is better determined in V, we chose 
to include both. We thus make two assumptions: 1) the stellar flux fraction is the same over the 450 to 850~nm wavelength range, 
and 2) the archival JOHNSON V-R color of 0.32~mag is not variable. Deviations from these assumptions are expected to be small 
because a single flux fraction fits the 550-850~nm range well and because the flux-variability is dominated by low-amplitude pulsations 
that do not affect the V-R color too much. The error is dominated by the absolute uncertainty on the interferometric
flux ratios, so our final confidence intervals should be robust. The fitted fluxes are listed in Table~\ref{table:stellarFluxes}.

\onltab{1}{
\begin{table}
\caption{The fluxes used in the ``stellar SED'' fit. For each filter, both the total flux and the stellar flux, i.e. the total flux corrected with the stellar flux fractions of 
Table~\ref{table:LITpro}, are listed.}             % title of Table
\label{table:stellarFluxes}      % is used to refer this table in the text
\centering                          % used for centering table
\begin{tabular}{c c c c}        % centered columns (4 columns)
\hline\hline                 % inserts double horizontal lines
Photometric filter & Total & Stellar & Error \\    % table heading 
\hline                        % inserts single horizontal line
   JOHNSON.V & 5.45\tablefootmark{a} & 6.00 & 0.15 \\      % inserting body of the table
   JOHNSON.R & 5.13 & 5.68 & 0.12 \\
   2MASS.H\tablefootmark{b} & 4.21 & 4.86 & 0.11 \\
   2MASS.Ks\tablefootmark{b} & 3.41 & 4.92 & 0.1 \\   
   JOHNSON.H & 4.39 & 5.04 & 0.12 \\
   JOHNSON.K & 3.35 & 4.85 & 0.12 \\
\hline                                   %inserts single line
\end{tabular}
\tablefoot{\tablefoottext{a}{Average of AAVSO light curve over 2011-2012},\tablefoottext{b}{Mt. Abu measurements.}}
\end{table}
}

We fit the SED with the \citet{1993yCatKurucz} model atmospheres. We use the grid-based method of \citet{2011AADegroote}, which allows us to identify correlations
between parameters and to take them into account in determining parameter uncertainties. It uses an $\chi^2$ statistic with five degrees of freedom to
determine the goodness of fit and confidence intervals (CI) for the five parameters. In practice, a model SED is corrected for interstellar reddening with
the reddening law of \citet{2004ASPCFitzpatrick}, then integrated over the required photometric passbands, and finally scaled to the measurements 
by optimizing the angular diameter.

As expected, T$_{\mathrm{eff}}$ and E(B-V) are correlated since 
reddening and effective temperature similarly affect the V-K color.
In the literature, different values for both parameters are quoted: based on photometry 
\citet{1990ApJLuck} adopt E(B-V)=0.1 and find T$_{\mathrm{eff}}$=6400~K, while \citet{1993AAWaters} find the best fit for E(B-V)=0.0 and T$_{\mathrm{eff}}$=6500~K. 
\citet{2011BaltAKipper} suggests an E(B-V)$>$0.1 based on the equivalent width of the interstellar component of the Na~I line. 
Spectroscopic effective temperatures
are consistently in the range 6500-6600~K \citep{1990ApJLuck,2011BaltAKipper}, except for the value of 7177~K found by \citet{2007PASJTakeda}. The latter study also found
a larger gravity ($\log$~g=1.66) and turbulent velocity (v$_t$=8~km/s). Finally, the Galactic extinction maps of \citet{1992AAArenou}, 
\citet{1998ApJSchlegel}, and \citet{2001ApJDrimmel} all give E(B-V)$\sim$0.1 for a distance larger than 1~kpc in the direction of 89\,Herculis. 
We adopt the spectroscopic temperature of 6550~K and
E(B-V)=0.07$\pm$0.04, both fully in agreement with our new SED. The derived CI are shown online in Figure~\ref{figure:chi2map-SED}.

The T$_{\mathrm{eff}}$ and $\theta_\star$, are not correlated strongly: the near-IR fluxes mainly delimit $\theta_\star$ since they are in the Rayleigh-Jeans regime. Our final angular diameter, at
T$_{\mathrm{eff}}$=6550~K, is $\theta_\star$=0.435$\pm$0.008~mas, which is perfectly within the range determined in Sect.~\ref{subsection:nearIRring} based
on the H band interferometric data. The final %reddened and original 
stellar SED models are shown in the left panel of Fig.~\ref{figure:SED}. 

\begin{figure}
\centering
   \includegraphics[width=8cm,height=7cm]{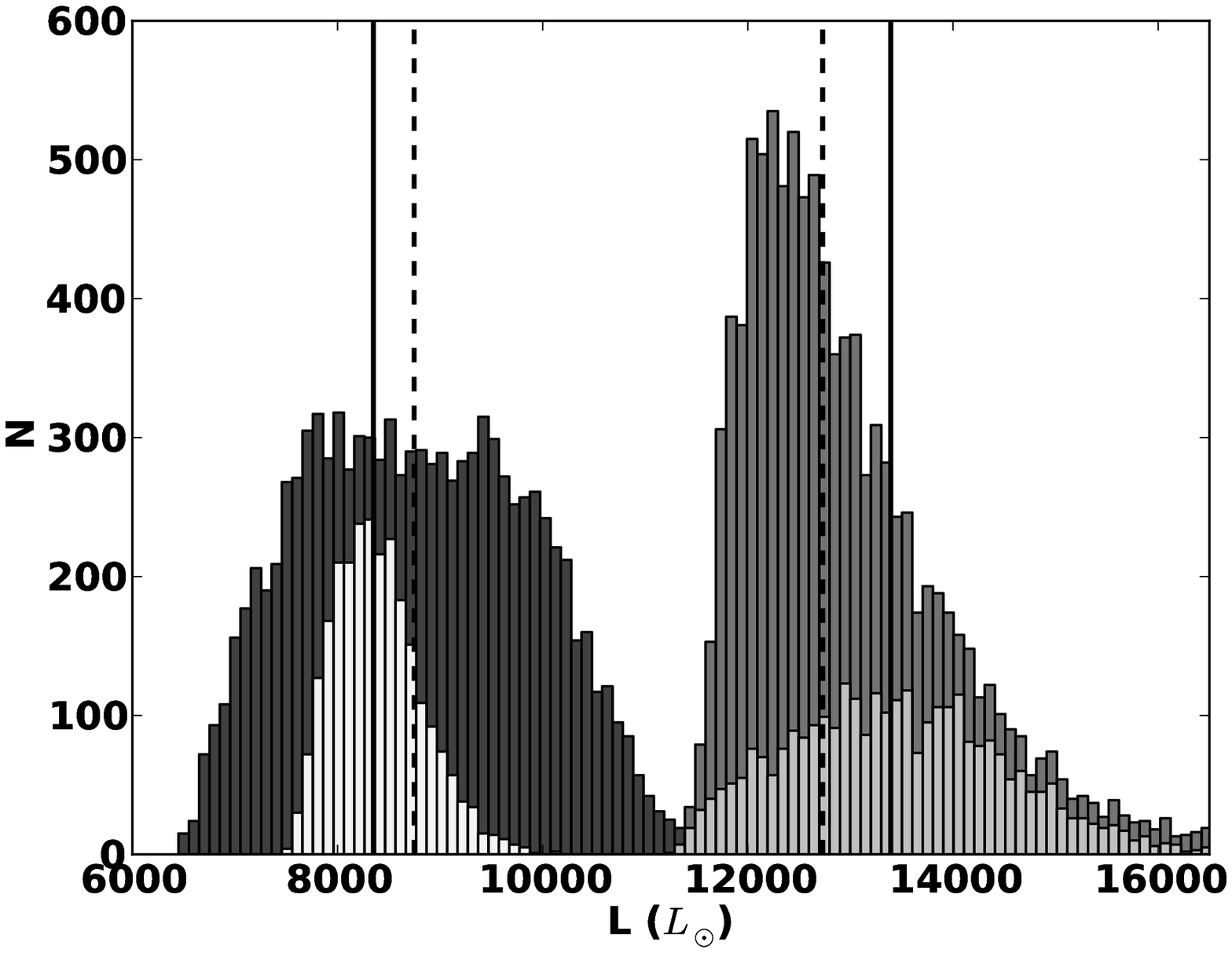}  
     \caption{Luminosity histogram of the stellar SED fit. The left and right histograms show the results of the fit to the flux ratio corrected and total photometry, 
     respectively. The full histogram (10000 points) is shown in dark gray, while the light gray
     subsamples were obtained thanks to the spectroscopic temperature constraint. The full and dashed lines are the median values of the light
     and dark histograms, respectively, and have values of 8350~and 8750~$L_{\odot}$ for our revised stellar photometry. The histograms 
     were made assuming a distance of 1.5~kpc.}
     \label{figure:Lumis}
\end{figure}

We made a stellar luminosity histogram by sampling the model parameter
space according to our derived probabilities.        %the probabilities derived from the SED fitting. 
Model bolometric fluxes are 
converted into luminosities by assuming a distance of 1.5~kpc. Following the measured parallax, this is the distance with the highest probability. Due to Lutz-Kelker bias
and the rather large error on the parallax, it is likely an underestimate of the true distance and hence of the derived luminosity. In general, our 
results are unaffected by this large uncertainty since the same distance is consistently used to convert the interferometric angular quantities 
into physical scales. By imposing the spectroscopic constraint on T$_{\mathrm{eff}}$, we adopt a final luminosity of 
8350~L$_\odot$, the median value of the subset of the histogram shown in Fig.~\ref{figure:Lumis}. 
As a comparison, we also show the histograms that would be obtained without our flux ratio correction 
(selecting only photometry up to J band). The median luminosity is then 13\,400~L$_\odot$.

\subsection{The circumstellar SED}\label{subsection:HERMES}
\begin{figure}
\centering
    \includegraphics[width=8cm,height=7cm]{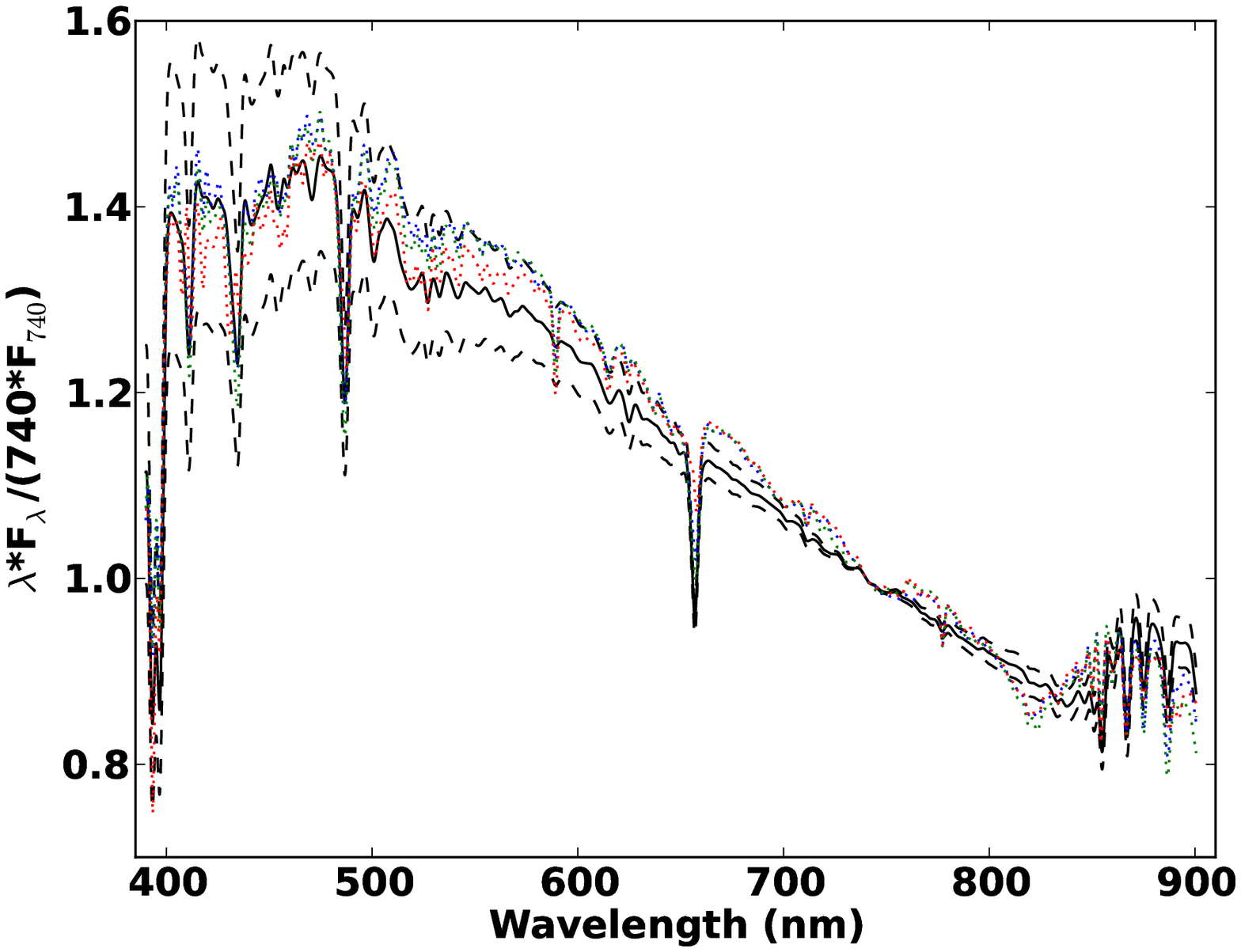}
     \caption{Hermes spectral shapes (dotted lines, colors denote the observation epoch) compared to spectral shapes from different 
     atmosphere models convolved to the same resolution. The full black line is the model shown in Fig.~\ref{figure:SED}. 
     The dashed lines have the same parameters, except for the reddening, which is E(B-V)=0.12 and 0.02 for the lower and upper line, respectively.}
     \label{figure:Hermes}
\end{figure}

The ``circumstellar SED'', obtained by subtracting the stellar SED from the total fluxes, is shown in the right panel of Fig.~\ref{figure:SED}, together 
with the stellar (but rescaled) Kurucz model. Fig.~\ref{figure:Hermes} contains the calibrated Hermes spectral shapes. 
Overplotted are three similarly convolved and scaled stellar Kurucz models with E(B-V)=0.02, 0.07, and 0.12 for the upper dashed, full, and 
lower dashed line, respectively. The observations follow the best-fit stellar model 
with E(B-V)=0.07 in both figures, although a careful inspection of the
Hermes spectral slopes shows that the measurements are slightly bluer
in the 500-700~nm range. This could be due to (the more abundant) small emission lines or
to a residual in the calibration. 
The Hermes spectrum is the sum of the stellar and circumstellar contribution. So the absence
of a color difference between the measured spectral shape and that of the star alone independently confirms that the process 
responsible for the circumstellar flux is essentially gray over the whole optical wavelength range. 

A basic quantity, often used as a proxy for the scaleheight of a passive dusty disk, is the ratio 
of the infrared excess over the stellar bolometric flux F$_{\mathrm{IR}}$/F$_\star$. \citet{2003AADominik} estimate the disk's half-opening angle $\alpha$
by approximating the disk as an optically thick cylinder that redistributes all received flux isotropically. The height of this cylinder 
determines which fraction of the stellar luminosity is intercepted. However, their derivation is only valid for small opening angles. 
Following their line of reasoning, one can more generally write
\begin{equation}
 F_{\mathrm{IR}} = \frac{L_\star}{4 \pi R^2} 8 \pi R^2 \sin^2 (\alpha/2) \frac{\pi D^2}{4 \pi d^2}
\end{equation}
and
\begin{equation}
 F_\star = \frac{L_\star}{4 \pi d^2} \pi D^2,
\end{equation}
with $L_\star$ the stellar luminosity, d the distance to the observer, and D the diameter of the observer's collecting area.   
Here, the fraction of the total solid angle subtended by the disk, as viewed from the star, is $8 \pi R^2 \sin^2 (\alpha/2)$, computed by integrating a sphere 
azimuthally and between polar angles $\alpha$ and $-\alpha$. 
This leads to the relation
\begin{equation}
 \frac{F_{\mathrm{IR}}}{F_\star} = 2 \sin^2 (\alpha/2)
\end{equation}
and a half-opening angle of $\alpha=67.5^\circ$ (see Fig.~\ref{figure:sketch}), obtained by integrating our revised stellar SED and the IR part of our circumstellar SED 
(F$_{\mathrm{IR}}$/F$_\star = 0.62$). If the large fraction of circumstellar scattered light is included as well, then we find a ratio of reprocessed
over stellar flux of F$_{\mathrm{RP}}$/F$_\star = 1.29$. In the isotropic approximation, F$_{\mathrm{RP}}$/F$_\star$ cannot become larger 
than one. Clearly, a significantly nonisotropic process must be at work in the 89 Her system.

% The reasoning is that it equals the fraction 
% of the stellar luminosity that is reprocessed/redirected by the disk. This translates into geometrical terms as the fractional solid 
% angle subtended by the disk. The latter, F$_{\mathrm{RP}}$/F$_\star \times 4 \pi R^2$, should be equal to the disk surface area, which is 
% that of a cylinder in a fully optically thick approximation, $2 \pi R \times 2 H$. We changed F$_{\mathrm{IR}}$ to F$_{\mathrm{RP}}$ since 
% it is the total amount of reprocessed light that matters, which is not necessarily equal to the infrared excess. 
% This leads to the relation:
% \begin{equation}
%  \tan \alpha = \frac{H}{R} = \frac{F_{\mathrm{RP}}}{F_\star},
% \end{equation}
% and half-opening angles of $\alpha=32^\circ$ and $\alpha=52^\circ$, obtained by integrating our revised stellar and circumstellar SEDs 
% (F$_{\mathrm{IR}}$/F$_\star = 0.62$ and F$_{\mathrm{RP}}$/F$_\star = 1.29$). 
% Note that we do not yet know whether all circumstellar flux is coming from the disk (see Sect.~\ref{section:discussion}).

Finally, we emphasize that our finding of an essentially gray circumstellar flux is
not fully independent from our previous assumptions. A wavelength-dependent flux ratio between 0.56
and 0.85~$\mu$m is possible since our observations are not very sensitive to this. Although our CI were determined conservatively 
as they are dominated by the error on the optical flux ratio, a significant wavelength dependence could result in an underestimated E(B-V) upper limit.
The angular diameter is not affected since our near-IR flux ratios are more robust. 
%This has implications for the value of F$_{\mathrm{RP}}$/F$_\star$ and the color of the circumstellar flux. 
Our F$_\star$ follows directly from $\theta_\star$ and 
the spectroscopic T$_{\mathrm{eff}}$ and is independent of the reddening, while F$_{\mathrm{RP}}$ depends on it twice. 
First, the circumstellar fluxes in the blue part of the SED are an extrapolation: they are obtained by subtracting 
the \textit{reddening-dependent} stellar Kurucz model from the measured (and still reddened) fluxes. 
Second, these reddened circumstellar fluxes need to be dereddened to obtain F$_{\mathrm{RP}}$. So our derived
F$_{\mathrm{RP}}$/F$_\star$ can still be underestimated, and the circumstellar flux would then be blue instead of gray.

%______________________________________________________________

\section{Discussion} \label{section:discussion}
In this study we present the first multiwavelength long-baseline interferometric data of a post-AGB binary with circumbinary disk that covers
wavelengths from 0.55 to 2.5~$\mu$m. 
Our most intriguing result is the discovery of a strong resolved optical flux component. 
We considered two geometric models to explain our observations: a binary and a star+ring, which both technically
fit our optical data equally well based on their $\chi^2_r$. Nevertheless, we argue that the binary model is hard to understand in terms of 
our near-IR data as well as from a physical point of view.
A star+ring model offers a more natural explanation for all our optical visibilities and is compatible in size with 
our near-IR findings. The optical ring can range from thin with a diameter of 5.5$\pm$1.0~mas (a circle) to extended 
with an outer diameter of 10$\pm$2~mas (a UD). Our analysis of the SED and the Hermes spectral slopes showed that the spectrum of the extended
emission is very similar to that of the post-AGB star, which strongly suggests a scattering process as its origin.

\begin{figure*}
\centering
    \includegraphics[width=18cm]{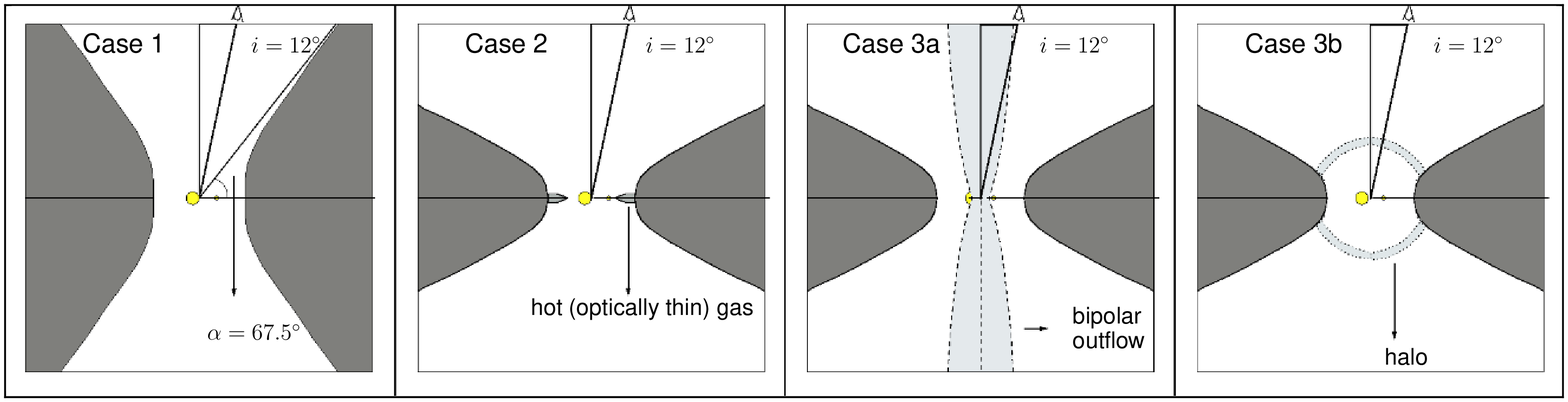}
     \caption{Pictographic representations of the different geometries that are discussed in Sect~\ref{section:discussion}.}
     \label{figure:sketch}
\end{figure*}

In the following we discuss three ways to explain our observations (see Fig.~\ref{figure:sketch}). Given the similarities 
between post-AGB and protoplanetary disks, we relate our results to both of them.

\subsection{Simply the circumbinary disk}
In the simplest case, all optical flux is light scattered off the inner rim of the dust disk. 
The question is whether a passive disk can reach the necessary scaleheight and redirect 
the required fraction of the stellar luminosity into our direction. Although estimated from very simple considerations, the 
large $\alpha$ of 67.5$^\circ$ shows that even to just account for the IR excess the disk already needs a large 
scaleheight. This could be favorable to account for the optical scattered light.
Asymmetric scattering is preferentially forward directed \citep[see, e.g.,][]{2013AAMulders}, so with the small inclination
of $\sim12^\circ$, our viewing angle onto the system is very unfavorable unless there is material high enough above the orbital plane. 
Quantitatively, this also depends on the properties of the dust, such as composition and grain size. 
However, it remains to be seen whether the optical and near-IR fluxes and visibilities can be made compatible 
in this geometry. A circumbinary disk with a very large scaleheight might result in a too thin, circle-like, near-IR emission geometry.
The optical visibilities are, however, compatible with the near side of the disk being brighter in scattered light, 
given the good fit of both a binary and a star+ring model to the NPOI observations.
Detailed radiative transfer modeling is required to tell whether this geometry is realistic.
Here we restrict ourselves to a qualitative discussion. % with respect to the literature.
 
Protoplanetary disks have been well studied in scattered-light images, especially with the HST
\citep[see, e.g.,][and references therein]{2013ApJGrady,2013ApJCox,2013AAMulders}, but also with Subaru 8.2m near-IR 
coronagraphic imaging \citep{2004ApJFukagawa,2006ApJFukagawa} and sparse aperture masking experiments or adaptive 
optics imaging on the VLT \citep{2013ApJCieza} or Keck \citep{2011ApJMcCabe,2001NaturTuthill}. 
These and other studies find integrated scattered light flux ratios of up to a few percent, while 
we find that 35-40\% of the total optical flux is resolved.
All the above methods have a limiting resolution of $\sim$30-50~mas, which corresponds to 5-10~AU for the nearest star-forming 
regions at 200~pc. For all but the brightest Herbig Be sources, these scales probe the bulk of the disk and not the inner rim. 
Near-IR interferometers do have the required resolution, and many inner rim regions have been resolved in this way (see further). However, 
at these wavelengths and with their typically sparse uv coverage, it is impossible to make a distinction between thermal and scattered light, contrary 
to the current study. To our knowledge, the only optical study of a protoplanetary disk and at a similar angular, but ten times higher spatial, resolution 
are the results on AB Aur with CHARA/VEGA \citep{2010AARousseletPerraut}. They resolved the H$\alpha$ line-forming region at sub-AU scales 
but found the continuum to be unresolved. \citet{2011AABonneau} did find a 15\%, fully resolved, optical continuum flux from a circumbinary dust 
reservoir in the interacting binary system $\nu$ Sgr, but it had rather large error bars. 

No published radiative transfer disk model, at viewing angles close to pole-on, has yet predicted such a large optical scattered flux as is found here. 
This implies that if we probe scattered light off the inner rim, the models are either lacking 
an important ingredient or post-AGB disks are different from protoplanetary ones or even 89\,Her is special in some way. 
In any case, evolved star disks are likely to be more puffed up than protoplanetary ones, 
given their high luminosities and small central masses.

\subsection{The circumbinary disk + an inner gas disk} 
Our results show a preference for an extended emission region, starting close to the central object.
Since dust only survives beyond the sublimation radius, some of this flux may be gaseous emission from within the inner rim. 
This is not unlikely given the detection of CO emission at 2.3~$\mu$m, needing temperatures $>$2000~K. 

Such gaseous emission was often suggested in the literature as the missing component 
to explain near-IR interferometric data of protoplanetary disks. \citet{2002ApJMonnier} found that certain high-luminosity sources are seriously undersized
with respect to their size-luminosity relation. This was later confirmed \citep{2005ApJMonnier,2004ApJEisner} and attributed 
to the presence of a (geometrically thin, but optically thick?) gaseous disk within the dust sublimation radius. 
More evidence was claimed in the studies of \citet{2005ApJAkeson} for some T Tauri stars 
and in the studies of \citet{2008ApJKraus} and \citet{2008ApJTannirkulam2} for some Herbig Ae/Be sources. Only the latter study had a 
sufficient spatial resolution to resolve the K band continuum emission from within the expected inner rim. Its authors attempted to empirically derive the sources 
of opacity responsible for the emission from the spectral shape of the flux-deficit between 1.25~and 10~$\mu$m. 
A slight preference for free-free and bound-free opacity of H$^-$ (5000~K) or neutral hydrogen (8000~K) was found over a mixture dominated by molecular opacity 
(2000-2500~K), which was due to the lack of strong molecular features between 4~and 10~$\mu$m. The H$^-$ and neutral hydrogen models overshoot their observations 
at wavelengths $\leq$2~$\mu$m and produce a significant optical flux. However, they did not investigate whether 
such extreme conditions are physically realistic. 

A more extensive data set on HD163296, one of the targets in \citet{2008ApJTannirkulam2}, was presented 
by \citet{2010AABenisty} and compared to simple gaseous disk models in a bit more detail. The optically thick models of \citet{2004ApJMuzerolle} 
were similarly rejected because of the absence of strong molecular emission features in the SED. Additionally, they presented SEDs of non-LTE gaseous components with
varying densities and temperatures and showed its spectral dependence to be incompatible with the observations. It might be tempting 
to interpret our resolved optical flux in terms of these hot gas models, but the latter argument also holds for 89\,Her. While H$^-$ is the foremost source of 
continuum opacity below 8000~K, bound-free processes dominate above it. In both cases the spectral dependence
is incompatible with our smooth circumstellar SED. The H$^-$ opacity reaches a maximum value 
at 850~nm and drops steeply shortwards.

Moreover, the source function needs a temperature $>$4500~K (assuming optically thick radiation) over the full emitting 
surface area to reproduce our visibilities and fluxes. Close to the illuminating source, the combination of 
such a temperature with a reasonable density (hence optical depth) \textit{might} be attainable. 
But due to flux dilution, the temperature drops quickly with radius as a power law with exponents of 1/2 and 3/4 for a flaring or flat reprocessing 
disk, respectively \citep{1987ApJKenyon}. Given the large size of the post-AGB star and our observations that do not support a bright companion, accretion 
energy cannot contribute significantly. Since the density also drops with radius, the disk becomes too optically 
thin at large radii very quickly. By decoupling the gas from the dust temperature, 
\citet{2009AAWoitke} found in their detailed thermochemical protoplanetary disk models a hot surface layer that bends around the inner 
rim with temperatures in excess of 4000~K. However, the density in this small region is too low to produce a significant continuum
optical depth and flux. On the other hand, \citet{2004ApJMuzerolle} found typical temperatures 
of only $\sim$1100~K (assuming LTE opacities), even close to the central star. 
All these models are very crude and as \citet{2010AABenisty} remarked, more realistic models that treat the 
transition from optically thin to optically thick gas layers in a dust-free environment are needed to definitely
rule out inner gas disks as a dominant effect. 

Alternatively, \citet{2010AABenisty} proposed an inner dusty disk consisting of highly refractory grains that manage to survive 
the low-pressure, high-temperature environment within the inner rim of the ``silicate-disk''. This model reproduced their near-IR interferometric 
data perfectly but required high temperatures of $\sim$2200~K at the inner radius. Using a more sophisticated treatment 
of dust sublimation and condensation physics within a Monte Carlo radiative transfer code, \citet{2009AAKama} also found 
solutions with a large optically thin region, covering up to 70\% of the surface inside the inner rim.
This geometry offers advantages as it allows for the existence of a large surface within the actual rim with possibly a distinct grain population. 
However, from a scattering perspective, the small geometric thickness of this region makes it energetically an unlikely option 
to account for our optical observations.

%%Their focus was on doing a theoretical parameter study of the inner rim physics, and not on predicting observables. 
%It is unclear what this optically thin region holds in terms of scattered flux at pole-on viewing angles. 
%%We don't know whether \citet{2010AABenisty} included scattering, 
%%but their ``inner disk SED'' looks like a perfect blackbody without any short-wavelength component, suggesting they either did not include it or used very small albedos. 

%But several questions come up as well. First, AB Aur was found by \citet{2008ApJTannirkulam2}
%to have such a smooth inner emission region in the near-IR while \citet{2010AARousseletPerraut} 
%did not resolve it in the optical continuum. Second, the light scattered away by this inner region cannot reach the 
%inner rim of the optically thick part of the disk, reducing its scaleheight and therefore compensating the whole effect. 

\subsection{The circumbinary disk + a (bipolar) outflow}  
The emission morphology might be more complex. The rather large errors on the NPOI visibilities prevent a 
firm conclusion on any position angle dependence of the visibility, but the good fit of the binary model shows that it cannot be excluded either. 
A bipolar outflow seen pole-on, originating as a stellar wind or jet, might naturally give rise to the 
presence of small-scale structure, including arc or knot-like features. There are several arguments in favor of this model. First, there is the large-scale 
hour-glass-like nebula discovered by \citet{2007AABujarrabal}, which must have an origin in the central object. 
Second, bipolar outflows are commonly inferred to explain high-resolution spectroscopic time series of many classes of evolved binaries, 
in particular post-AGB ones like the Red Rectangle \citep{2009ApJWitt,2011MNRASThomas,2013MNRASThomas} or BD+46$^\circ$442 \citep{2012AAGorlova}. 
The weak, neutral or low-excitation-level emission lines of metals (e.g. Fe I near 805~nm) were already interpreted by \citet{1993AAWaters} 
in terms of a collisionally excited interaction between a modest stellar wind and the circumbinary disk. The same lines, which are variable 
in strength but not in velocity, are seen in emission in QY Sge \citep{2002MNRASKameswara} and BD+46$^\circ$442 \citep{2012AAGorlova}, 
but with a different width that the latter authors suggest to be an inclination effect. Also, the weak and variable H$\alpha$ P Cygni profile 
is a clear indication for a small mass loss rate of $\sim$10$^{-8}$~M$_\odot$~yr$^{-1}$ \citep{1969ASSLSargent}. The main question in 
this outflow scheme is why the scattered light is so well confined to the size of the disk inner rim (compare the panels in Fig.~\ref{figure:chi2maps1}).

Jets or bipolar outflows are often detected around YSOs for a wide range of central masses, and their presence is strongly 
linked to accretion phenomena \citep{1999NewARKonigl}. These jets are detected over the full electromagnetic spectrum, 
most notably through maser, free-free, and synchrotron emission at radio wavelengths, 
IR molecular tracers, and in shock-excited near-IR/optical/UV spectral lines \citep{2007ApSSBally}. %However, the highly collimated
%and supersonic jets are not seen in optical scattered light once the material from their parental cloud has been cleared, while they would be resolvable. 
For highly obscured objects, significant bipolar reflection nebulae are seen and interpreted as scattering by dusty material within or on the walls of
the outflow cavities produced by the jet \citep{1999AJPadgett}. 

A similar explanation (but with a different origin for the dust) has been given for the 
HST images of bipolar planetary (PNe) and proto-planetary nebulae (PPNe) in scattered light 
\citep{1998ApJKwok,2000ApJKwok,2000ApJUeta,2002AABujarrabal,2004AJCohen,2008ApJSiodmiak}. Post-AGB binaries are classified as ``stellar'' 
in these HST surveys, which is corroborated by the small scale of the optical circumstellar flux found here.

% \citet{2011ApJKoning} in particular present a simple, but convincing, geometric model 
% for the shaping of the Red Rectangle nebula based on a spherical halo, a bi-conical cavity and a central illuminating source. 
% A key advantage of their model they claim, is that it explains the unique morphology of the Red Rectangle in terms of a rather universal bipolar model, that is 
% applicable to many PPN (or in a way even to PN). 

Based on optical and UV spectropolarimetry, \citet{1987AAJoshi} claim the presence of two geometrically distinct
dust populations around HR4049, which consist of differently sized particles. Later, \citet{1999MNRASJohnson} confirm this result and associate the component
with the smallest particles ($\leq0.05 \mu$m), seen mainly at wavelengths $<$2000~\AA, with a bipolar structure. \citet{1987AAJoshi}
found for 89\,Her an even larger degree of polarization than for HR4049, but without the latter's PA dependence. 

% Observationally, no clear evolutionary link has been established between PNe, PPNe and post-AGB 
% binaries \citep{2009PASPDeMarco}. The latter are typically classified as ``stellar'' in HST surveys searching for large-scale nebulosity 
% \citep{2000ApJUeta,2008ApJSiodmiak}. The small scales found in our resolved optical data, corroborate that there is no evidence for the presence of 
% a large-scale outflow in scattered light around 89\,Her, unless it is very well collimated. 

The main advantage of an outflow is that it allows having material at a high altitude above the disk midplane that can efficiently 
(forward-)scatter light into our
line of sight that would otherwise freely escape. Although contested, optically thin halos are an alternative way to do this and are often inferred to explain 
(near-IR resolved) data of protoplanetary disks \citep{2012AAChen,2011AAVerhoeff,2010AAMulders}. Explanations for halo formation range 
from collisions between planetesimals with highly inclined orbits \citep{2011AAKrijt} to dust entrainment 
in upper disk layers, where dust and gas are not thermally coupled \citep{2009AAWoitke} and the gas thus has a higher vertical extension \citep{2011AAVerhoeff}. 
In any case, ``halos'' are likely too optically thin if they are dynamically stable. In the model of \citet{2011AAVerhoeff}, 
the halo contributes at most 10\% of the stellar optical flux. 

%Optically thin halos are an alternative way for this and are often inferred to explain 
%(near-IR resolved) data of protoplanetary disks \citep{2012AAChen,2011AAVerhoeff,2010AAMulders}. Although contested, 
%A fully spherical halo as well as a bipolar outflow, puts material in the direct line of sight, increasing the extinction. An E(B-V) twice as 
%large is still within our CI, but would imply an even larger fraction of scattered flux in the blue part of the SED. A vertical extension 
%of the circumbinary disk, like in the second ``halo-forming scenario'', might circumvent this problem.  

%______________________________________________________________

\section{Conclusions}
Our optical interferometric data show that one should be careful not to overlook a possible scattering component when establishing the energy budget 
in systems with significant circumstellar material as derived from the IR excess. By separating the direct stellar light from the reprocessed, 
circumstellar light between 0.5 and 2.2~$\mu$m, we reassessed the
stellar and circumstellar spectral energy distribution and revised the stellar luminosity of 89\,Her from 13\,400 to 8350~L$_\odot$, assuming a distance of 1.5~kpc. 
This lower luminosity corresponds to a decrease in angular (and thus
also physical) diameter from the previously assumed $\sim$0.65 to the
directly measured 0.435$\pm$0.008~mas. 

The circumstellar luminosity is then likewise increased, resulting in a ratio of F$_{\mathrm{IR}}$/F$_\star = 0.62$ and F$_{\mathrm{RP}}$/F$_\star = 1.29$. 
The former leads to a large half-opening angle of the circumbinary disk of 67.5$^\circ$ 
\textit{if} all the observed IR circumstellar flux is reprocessed light by the disk. The latter, on the other hand, requires a nonisotropic process.
In Paper II we will use a radiative transfer code to test whether a passive circumbinary disk can be created that reproduces our observations.

Alternatively, adding geometric components could ease the requirements on the scaleheight of the disk, and we
discussed several possibilities. A bipolar outflow is an interesting option from an evolutionary perspective but requires explanation in terms
of projected size. A vertical extension of the disk into a kind of halo is also possible, but might be too optically thin. 
An inner gas disk is deemed unlikely, but definite exclusion of it requires more realistic models in non-LTE conditions. 
%The latter is in any case compatible with the K band emission geometry, for which we find a rather extended region starting at an inner radius of 
%at most twice the orbital separation from the center-of-mass. Refractory dust that can survive high temperatures could be an alternative. 
%The rather mild depletion pattern seen in 89\,Her's post-AGB photosphere does not favor either option, as there are several ways to explain it.

Our redefined SEDs show that the circumstellar energy budget is already dominated by scattered light in H band. 
%Maybe the dissimilarity between our AMBER epochs is related to this, but we prefer the interpretation of the 2007 data as being erroneous. 
Compared to the K band, we find a slight preference for 
a smaller (projected) emission region in H band and a clear preference in the optical, which could be a hint for the presence of 
material above the orbital plane that mainly contributes through scattering. 

An important consequence of our findings is that determining luminosities of post-AGB binaries from SED fitting is even more complex 
than previously assumed. On top of the difficulties related to inclination, reddening, and distance (still the dominant source of uncertainty), 
one now also has to take into account a possibly dominant scattered light contribution. 
Its exact origin decides whether 89\,Her is an extreme or average case 
and what will be the spread in scattered light fractions for other objects. 
If resolved observations are the only way to detect this extended optical emission, a detailed comparison of post-AGB objects
with evolutionary tracks might become problematic because LMC sources cannot be resolved spatially and distances to galactic sources will remain 
uncertain, at least until the Gaia satellite comes online.

%The impact of our detection on other disk sources is difficult to assess, as it depends on the physical origin of the scattered light. 
%If simply the disk or a kind of halo is responsible, similar detections might be expected for protoplanetary disks. 

%In the future, more interferometric observations at optical, near-IR and also mid-IR 
%wavelengths are highly needed to better constrain the properties of the close circumstellar environment of these interacting 
%binary systems. Many questions remain about the evolution of the disk, and how it interacts with the central binary and vice versa. 

\begin{acknowledgements}
M.H. wishes to thank B. Acke and C. Dominik for the useful discussions about this work and P. Degroote for his SED-fitting scripts.
The Navy Precision Optical Interferometer is a joint project of the Naval Research Laboratory and the US Naval Observatory, in cooperation with Lowell 
Observatory, and is funded by the Office of Naval Research and the Oceanographer of the Navy. The authors would like to thank Jim Benson and the NPOI
observational support staff, whose efforts made this project possible. The CHARA Array is funded by the National Science Foundation through NSF grant AST-0606958,
by Georgia State University through the College of Arts and Sciences, and by the W.M. Keck Foundation.
The Palomar Testbed Interferometer was operated by the NASA Exoplanet Science Institute 
and the PTI collaboration. It was developed by the Jet Propoulsion Laboratory, California Institute of Technology, with funding 
provided from the National Aeronautics and Space Administration. This work has made use of services produced by the NASA Exoplanet Science Institute 
at the California Institute of Technology. The IOTA data would not have been possible without contributions from Mark Swain, 
Ettore Pedretti, Wes Traub, J.-P. Berger, and Rafael Millan-Gabet. We also thank SAO, U. Mass, NSF AST-0138303, NSF AST-0352723, 
and NASA NNG05G1180G for supporting IOTA development and operations.

%We acknowledge with thanks the variable star observations from the AAVSO International Database contributed by observers worldwide and used in this research.
\end{acknowledgements}

\bibliographystyle{aa}
\bibliography{aa_89Her}

\end{document}